\documentclass[12pt, preprint]{aastex}

\shorttitle{Modeling Emission in a Post-CME Current Sheet}
\shortauthors{Ko et al.}

\newcommand{\Ro}{R$_\odot$}
\newcommand{\Ros}{R$_\odot$~}
\newcommand{\lsim}{\lower.5ex\hbox{$\; \buildrel < \over \sim \;$}}
\newcommand{\gsim}{\lower.5ex\hbox{$\; \buildrel > \over \sim \;$}}

\usepackage{graphicx}
\usepackage{epstopdf}
\begin{document}

\title{Modeling UV and X-Ray Emission in a Post-CME Current Sheet}

\author{Yuan-Kuen Ko\altaffilmark{1}, John C. Raymond\altaffilmark{2}, Bojan Vr\~snak\altaffilmark{3}, Eugen Vuji\'c\altaffilmark{4}}
\altaffiltext{1}{Space Science Division, Naval Research Laboratory, Washington, DC 20375, USA}
\altaffiltext{2}{Harvard-Smithsonian Center for Astrophysics, Cambridge, MA 02138, USA} 
\altaffiltext{3}{Hvar Observatory, Faculty of Geodesy, Zagreb, Croatia} 
\altaffiltext{4}{Faculty of Science, Geophysical Department, Croatia} 

\email{yko@ssd5.nrl.navy.mil}

\begin{abstract}
 
A post-CME current sheet (CS) is a common feature developed behind an erupting flux rope in CME models. Observationally, white light observations have recorded many occurrences of a thin ray appearing behind a CME eruption that closely resembles a post-CME CS in its spatial correspondence and morphology. UV and X-ray observations further strengthen this interpretation by the observations of high temperature emission at locations consistent with model predictions. The next question then becomes whether the properties inside a post-CME current sheet predicted by a model agree with observed properties. In this work, we assume that the post-CME CS is a consequence of Petschek-like reconnection and that the observed ray-like structure is bounded by a pair of slow mode shocks developed from the reconnection site. We perform time-dependent ionization calculations and model the UV line emission. We find that such a model is consistent with SOHO/UVCS observations of the post-CME CS. The change of Fe XVIII emission in one event implies an inflow speed of $\sim$10 km/s and a corresponding reconnection rate of $M_A \sim 0.01$. We calculate the expected X-ray emission for comparison with X-ray observations by Hinode/XRT, as well as the ionic charge states as would be measured in-situ at 1 AU. We find that the predicted count rate for Hinode/XRT agree with what was observed in a post-CME CS on April 9, 2008, and the predicted ionic charge states are consistent with high ionization states commonly measured in the interplanetary CMEs. The model results depend strongly on the physical parameters in the ambient corona, namely the coronal magnetic field, the electron density and temperature during the CME event. It is crucial to obtain these ambient coronal parameters and as many facets of the CS properties as possible by observational means so that the post-CME current sheet models can be scrutinized more effectively.

\end{abstract}


\section{Introduction}

Most models for coronal mass ejections (CMEs), regardless of what initiates the CME, predict a current sheet (CS) that develops beneath an erupting flux rope due to the stretching of the overlying coronal field (e.g., Lin \& Forbes 2000, Linker et al. 2003, Lynch et al. 2004, MacNeice et al. 2004, Manchester et al. 2004). Magnetic reconnection in the CS reduces tension restraining the outgoing flux rope and at the same time produces post-CME loops beneath the reconnection point. Outflows along the CS and a temperature higher than that in the ambient corona are expected inside the CS as magnetic energy is converted to kinetic and thermal energy due to reconnection.
This standard flare-CME picture (Figure 1a) is supported by observations such as loop-top hard X-ray sources (e.g. Masuda et al. 1994), upward growth of flare loops (e.g.  \v{S}vestka 1996), separating flare ribbons, and hotter post-flare loops lying higher than the cooler post-flare loops (e.g. van Driel-Gesztelyi et al. 1997). However, few observational signatures of the CS {\it above or near} the X-point were reported until recently. Sui et al. (2003,2004) reported signatures of a CS at both sides of the reconnection site from X-ray observations by the Reuven Ramaty High Energy Solar Spectroscopy Imager (RHESSI). Savage et al. (2010) reported a post-CME CS feature observed by Hinode/XRT with downflows/upflows that allow the X-point to be located. Innes et al. (2003b) and Wang et al. (2007) reported high speed outflows in opposite directions away from the reconnection site based on Doppler shift signatures of Fe XIX and Fe XXI lines observed by Solar Ultraviolet Measurements of Emitted Radiation (SUMER) on the Solar and Heliospheric Observatory (SOHO). White light (WL) observations of CME events by Solar Maximum Mission (SMM) and Large Angle Spectrometric Coronagraph (LASCO) on SOHO have recorded many occurrences of a thin ray appearing behind a CME eruption (we will call it `WL ray' hereafter) that closely resembles a post-CME CS in its spatial correspondence and morphology (e.g. Webb et al. 2003, Ko et al. 2003, Lin et al. 2005). Observations by Ultraviolet Coronagraph Spectrometer (UVCS) on SOHO of high-temperature emission (3-6 million degrees) that lies along such WL ray in between the erupting CME and post-CME loops provide strong support for the CS interpretation, as opposed to being just a usual streamer seen edge-on (Ciaravella et al. 2002, Ko et al. 2003, Bemporad et al. 2006, Lee et al. 2006, Ciaravella \& Raymond 2008). Figure 2 shows two examples of such events.

Even though there is strong observational support for the observed WL ray being the post-CME CS, that interpretation remains under scrutiny. One major cause of skepticism arises from the observed thickness of the WL ray, as well as the spatial extent of the high-temperature emission observed by UVCS, which are  much larger than what most reconnection models predict. The observed thickness is of the order of $10^4-10^5$ km (Lin et al. 2007,2009, Vr\v{s}nak et al. 2009) which is orders of magnitude larger than theoretical estimates of the thickness for the diffusion region and Sweet-Parker CS (see Bemporad 2008, Lin et al. 2009) even for the largest estimates of anomalous resistivity currently available, and taking projection effects into account. Bemporad (2008) and Lin et al. (2009) discussed several reconnection schemes with turbulence or instability (see references therein) that may produce CS thickness comparable to what was observed, such as turbulence and stochastic/fractal/time-dependent Petcheck-type reconnection. Model calculations have shown that the effect of thermal conduction can create a thermal `halo' around the current layer (Yokoyama \& Shibata 1997, 2001; Seaton \& Forbes 2009). This would contribute to the extent of the high-temperature emission observed by UVCS, although not for the WL emission (which depends mainly on the electron density). The electric potential
of the slow mode shocks might inhibit transport of energy into the upstream flow, reducing
this effect. To see which theories/models are more consistent with or to reject the CS interpretation, the theories or models for the CS must predict some physical quantities that can be tested by these observations. 

We should not forget that there is another piece of observational evidence that can be used to scrutinize the CS interpretation and constrain viable theories/models. As mentioned above, SOHO/UVCS has observed several CME events exhibiting high-temperature emission from the [Fe XVIII] $\lambda$974 line (formation temperature of 6 million degrees) at heliocentric heights of 1.5-1.7 \Ros that lay along the line connecting the erupting CME and the associated post-CME loops (e.g. see examples in Fig.2). Its location and the timing of its brightening following the eruption strongly favor the interpretation of a post-CME CS. It is then important to go one step beyond morphology and find out if a given CS model can predict physical quantities, such as density and line intensities, that agree with the observations. And if so, how the observed quantities would imply the physical conditions for the associated reconnection within the context of a given model. 

It is important to note that the analysis of emission lines
from current sheets has so far assumed that the plasma is
in ionization equilibrium, but this can be a poor assumption.
For instance, the ionization time scale for Fe XVIII is on
the order of $10^{10}$/$n_e$ seconds, or nearly 1000 seconds
for the densities estimated from UVCS observations, while the
flow time in the current sheet should be on the order of
a few tenths of a solar radius divided by an ouflow speed
on the order of 1000 $\rm km~s^{-1}$, which can be several
times smaller. It is therefore important to consider time-dependent ionization when interpreting emission line intensities.

Vr\v{s}nak et al. (2009) proposed a working hypothesis for the observed post-CME CS (i.e. WL ray), and provided the first attempt to calculate the electron density within the CS, and compare with the observations. This model assumes a steady state Petschek-like reconnection scheme (Petschek 1964). The reconnection takes place in the diffusion region (DR) and a pair of standing slow-mode shocks (SMSs) develop and extend out from the DR. At the SMS crossing, the inflowing coronal material is compressed and heated in a way that is mainly determined by the external plasma $\beta$, and the velocity of the outflowing jet within the region bounded by the SMSs is approximately equal to the external Alfv\'en speed (Aurass et al. 2002). They found that the observed electron density and morphology of the post-CME CS (including thickness) can be successfully explained by their model. In this paper, we adopt their model and calculate, for the first time, the expected UV and X-ray emission inside the CS. Strictly speaking, within the context of this model, the so-called post-CME CS seen as a ray-like structure in white light observations is actually the region bounded by the SMSs. This region is not a 'current sheet' per se, and the electric currents are concentrated only in the DR and SMSs (Vr\v{s}nak et al. 2009). Note that, however, we will use `post-CME CS', `CS',  or `WL ray' interchangeably throughout this paper to stand for this region.

Section 2 describes the models in detail. We adopt two approaches. One is the `Fully-mixed' model which is the direct adaptation of the model laid out in Vr\v{s}nak et al. (2009). The other is the `Streamline model' which assumes that the outflow jets out of the SMS crossings at different locations do not mix with each other. We describe the methods for calculating the electron density, electron temperature and time-dependent ionic fractions within the CS for both models. Section 3 presents the predicted UV and X-ray emission, and the expected frozen-in charge states as would be measured in-situ at 1 AU. We discuss the results in Section 4.

\section{Modeling the Post-CME Current Sheet}

Figure 1b illustrates the idea of the model of Vr\v{s}nak et al. (2009) that is based on the steady-state Petschek reconnection scheme. As the eruption stretches the oppositely directed coronal field beneath the flux rope, reconnection takes place at the DR under suitable conditions. A pair of standing SMSs then develops out of the DR both above and below. In this paper, we will concentrate on the region bounded by the pair of SMSs {\it above} the DR in this vertical CS configuration. Coronal material that crosses the SMS is compressed, heated, accelerated, and forms an outflow jet above the DR. The extent of compression and heating depends on the ambient (i.e. coronal) plasma $\beta$ at the location of the SMS crossing as in Aurass et al. (2002):
 \begin{equation}
{n_2\over n_1}= {{5(1+\beta)}\over{2+5\beta}}
\end{equation}
and
 \begin{equation}
{T_2\over T_1}= 1+{2\over{5\beta}}
\end{equation}
where $n_1$,$T_1$ and  $n_2$,$T_2$ are the electron density and electron temperature in the ambient corona and after the SMS crossing, respectively. $\beta$ is the ratio of gas to magnetic pressure in the ambient corona for fully ionized plasma (Mann et al. 1999):
\begin{equation}
\beta= {{1.92n_1kT_1}\over{(B_{cor}^2/8\pi)}}
\end{equation}
where $B_{cor}$ is the ambient coronal magnetic field at the SMS crossing. The outflow speed out of the SMS crossing is approximated to be the Alf\'ven speed for the inflowing material:
\begin{equation}
v_A={ {B_{cor}}\over{\sqrt{4\pi \times 1.17 n_1m_p}}}
\end{equation}

To calculate the physical properties inside this post-CME CS, one needs to take into account all outflows from the SMS crossings along the SMS (Fig.1b). At this point, we will investigate two models. One is the 'Fully-mixed' model (`Model 1', Fig.1c) in which plasma that just comes inside the CS through SMS mixes thoroughly with material that comes from below, as in Vr\v{s}nak et al. (2009). This is the `fluid' approach and the electron conductivity is large across the CS so that thermal equilibrium is achieved between the incoming and outflow material at a given height. The other is an extreme opposite of the first model, the 'streamline' model (`Model 2', cp. Fig.1b), in which plasma that just comes inside the CS through SMS forms its own `streamline' and does not mix with the material that comes in at other SMS crossings. This would be the case if certain conditions inside the CS, e.g. turbulence, prevent these flows (i.e. streamlines) from efficiently changing the thermodynamical properties of each other. We discuss the two models in more detail below.

Both models adopt the same ambient coronal conditions. As shown above, the properties inside the CS depend on the ambient plasma $\beta$, which in turn depends on the electron density, electron temperature and magnetic field profiles in the ambient corona. For the coronal magnetic field, we adopt the empirical model by Dulk \& McLean (1978):
\begin{equation}
B_{cor}(r) = 0.5(r-1)^{-1.5} \rm{G}
\end{equation}
where $r$ is the heliocentric distance from the Sun in \Ro.
Strictly speaking, this formula is applicable for approximately $1.02 \le r \le 10$ \Ros but we will use this formula up to 20 \Ro~(see also Vr\v{s}nak et al. 2002), the upper boundary of our calculations. The accuracy of this formula between 10 and 20 \Ros is not a concern for our study since, as we will see below, 1) the emission calculated here only have observational data available at locations much lower than 10 \Ros (because the emission drops with $n_e^2$), and 2) the change of time-dependent ionic fractions occur mostly below 10 \Ro. This formula gives a value of $1.6\times 10^{-4}$ Gauss (16 nT) at 1 AU which is also reasonable (e.g. see ACE L3 summary plots at the ACE Science Center, http://www.srl.caltech.edu/ACE/ASC/index.html).
To investigate the dependence on different coronal electron density and temperature profiles, we use two $n_{cor}$ and two $T_{cor}$ profiles. For the coronal electron density profiles, one (denoted as `N1') is a 'hybrid' profile which, for $r \le 5.66$ \Ro, is obtained for a streamer during the SPARTAN 201-1 mission (Guhathakurta \& Fisher 1995). Beyond 5.66 \Ro, we linearly interpolate between the measured values at 5.66 \Ros and a value of 211.2 cm$^{-3}$ at 43 \Ros (0.2 AU) which is based on Mann et al. (1999):
\begin{equation}
n_{cor}(r)=6.53\left({r\over{215}}\right)^{-2.16} \phantom{..} \rm{cm^{-3}}. 
\end{equation}
for  0.2 AU $\le r \le$ 5 AU.
The second $n_e(r)$ (denoted as `N2') is adopted from LeBlanc et al. (1998) which is of an analytic form of 
\begin{equation}
n_{cor}(r)=3.3\times 10^5 r^{-2}+4.1\times 10^6 r^{-4}+8.0\times 10^7 r^{-6}\phantom{..} \rm{cm^{-3}.}
\end{equation}
Note that both profiles are based on observations. For the coronal electron temperature profiles, we assume a general form (in K):
\begin{equation}
T_{cor}(r)= \cases {T_0 &for 1 \Ro $\le$ r $\le$ 2  \Ro; \cr T_0/(0.5r)^{0.92} &for r $>$ 2  \Ro. \cr}
\end{equation}
The index 0.92 is based on Totten \& Freeman (1995) for a spherically expanding wind. For $T_0$, we will compare results for $T_0=10^6$ K (`T1') and $2\times 10^6$ K (`T2').
The solar wind speed profile is then taken as:
\begin{equation}
v_{sw}(r)=2\times 10^8 {1\over n_{cor}(r)}\left({215\over r}\right)^2 \rm{cm/s}.
\end{equation}
based on the continuity equation $nvr^2=$constant with $nv=2\times 10^8$ cm$^{-2}$s$^{-1}$ at 1 AU.

Figure 3 plots the adopted $B_{cor}$, $n_{cor}$, and $T_{cor}$ profiles, and the corresponding plasma $\beta$. Also plotted (`1MK\_Parker') are those adopted by Vr\v{s}nak et al. (2009) using a Parker wind model (Mann et al. 1999) with 1 MK isothermal corona, and the same $B_{cor}(r)$ (Dulk \& McLean 1978). Note that none of the actual coronal parameters in the immediate vicinity of the SMSs are known for certain, and the coronal conditions are likely to vary from one event to another. Furthermore, Vr\v{s}nak et al. (2009) found that the coronal density is depleted from the pre-CME corona in the vicinity of the WL ray. In any case, we believe that these coronal profiles adopted here are reasonable and can be taken as generic choices. Nonetheless, it is necessary to investigate the effect of different coronal parameters on the calculated CS properties. Therefore we calculate the model results for three cases: 1) N1+T1 (denoted as `N1T1'), 2) N1+T2 (denoted as `N1T2'), and 3) N2+T2 (denoted as `N2T2') while keeping the same $B_{cor}(r)$. Cases `N1T1' and `N1T2' are used to investigate effects from different coronal temperature profiles. Cases `N1T2' and `N2T2' are used to investigate effects from different coronal density profiles. We take N1T2 as the reference case for comparison among different cases and models. The lower right panel of Fig.3 plots the plasma $\beta$, which determines the jump condition across the SMS (Eqs.(1) and (2)), for these 3 cases. The $\beta$ in Vr\v{s}nak et al. (2009, `1MK\_Parker') is also plotted for comparison.

\subsection{Model 1: The Fully-mixed Model}

Model 1 (Fig.1c) follows the same thermodynamic calculations as Vr\v{s}nak et al. (2009), except that we calculate the CS properties with three cases of coronal parameters described above. Readers are referred to the Appendix in Vr\v{s}nak et al. (2009) for a complete description of the model. We only briefly describe the concept here. For a given cell ($\Delta R$ in Fig.1c) at a given height $R_i$ above the DR, plasma comes into the CS through the SMS with density $n_2$ (Eq.(1)) into a given volume ($V_{SMS}$) that is determined by the ambient coronal parameters. There is also material coming from the cell below into a volume ($V_{thru}$) determined by spherical expansion. In this model, the outflow (i.e. exhaust flow) speed is taken as the local value of $v_A+v_{sw}$ (for details see Vr\v{s}nak et al. 2009) but $v_{sw}$ is almost always negligible in the considered height range. Under the assumption that the `incoming' and `through' material thoroughly mix with each other, the electron density flowing out of this cell into the cell above it can be calculated from mass conservation, along with the CS geometry. The electron temperature in the cell is calculated to be the average of the temperature after the SMS crossing (i.e. $T_2$, Eq.(2)) and that coming from the cell below under adiabatic cooling, weighted by the mass inside $V_{SMS}$ and $V_{thru}$ respectively. The density and temperature profiles along the CS (above the DR) can then be calculated iteratively, and they {\it depend on the ambient coronal parameters and the location of the DR}. Since the magnetic field in this configuration is perpendicular to the exhaust flow, the thermal conduction in the flow direction will be inhibited. Note that the inflow speed $v_{in}$, which is a factor that governs the CS geometry, is calculated from $rv_{in}B_{cor}=$constant under the steady state assumption and with $v_{in}=0.01 v_A$ at the DR. 

Figure 4 plots the resulting $n_e(r)$, $T_e(r)$ and $v_{out}(r)$ for the 3 cases of the coronal $n_e/T_e$ profiles, and for DR at 1.1 and 1.5 \Ros for each case. The profiles from Vr\v{s}nak et al. (2009) (`1MK\_Parker') and the electron densities derived from UVCS and LASCO data are also plotted for comparison. The profiles with DR heights in between 1.1 and 1.5 \Ros lie between the two curves for each case respectively. Note that models N1T1 and N1T2 have almost the same profiles in the CS because when beta is small as in these cases, the jump condition for $n_e$ is about the same (Eq.(1)). The factor of 2 difference in $\beta$ is compensated by the factor of 2 in $T_1$ (Eq.(2) implies $T_2/T_1 \sim \beta^{-1}$ for small $\beta$), leading to almost the same temperature $T_2$ in the CS. 
The electron density in the CS is higher than that in the ambient corona at all heights (comparing Figs.3 and 4, see also Eq.(1)), therefore the CS structure can stand out against the ambient corona as observed. 

In order to calculate the emission from the CS, we need to first calculate the ion charge states (i.e. ionic fractions) of the ions that contribute to the emission. With large outflows in the CS, the ion charge states do not always maintain ionization equilibrium when the electron density is low enough. The evolution of the ion charge states with the flow depends on the electron density and temperature profiles as well as the ion outflow profiles. For a given element, the evolution of ion charge states with the flow is expressed as (under steady state assumption):
\begin{equation}
{d y_{q}\over dt} = n_e\left(C_{i,q-1}y_{q-1}-C_{i,q}y_{q}\right)+
n_e\left(R_{rr,q+1} +R_{dr,q+1}\right)y_{q+1}-
n_e\left(R_{rr,q}+ R_{dr,q}\right)y_{q}
\end{equation}
where $r$ is the heliocentric height of the fluid element along the CS, $y_q$ is the ionic fraction of charge state $q$. $C_{i,q}, R_{rr,q}, R_{dr,q}$ (which mainly depend on $T_e$)
are the rates for electron impact ionization (including auto-ionization), radiative recombination and dielectronic
recombination respectively, out of the charge state $q$. The ionization and recombination rates are calculated using the most recent compilation by Bryans et al. (2006,2009 and references within). Here we assume that all ions have the same outflow speed. For each element, Eq.(10) for all charge states (i.e. Z+1 equations for an element with atomic number Z) are solved simultaneously. We calculate the ionic fractions for 13 most abundant elements: H, He, C, N, O, Ne, Mg, Si, S, Ar, Ca, Fe, Ni.
Note that the ions that just cross the SMS should still carry the charge state distribution at the ambient coronal temperature (even though the electron temperature jumps to $T_2$ after the crossing), and those calculated by Eq.(10) are for ions coming from the cell below (cp. Fig.1c) following the flow. Therefore, the charge state distribution in a given cell would be the average of the two regions weighted by the mass inside $V_{SMS}$ and $V_{thru}$, respectively, for that given cell. 

Figure 5 plots the resulting $y_q(r)$ for case N1T2 with DR at 1.2 \Ros for a few ions, along with the values in the case of ionization equilibrium. We can see that these ion charge states are far from ionization equilibrium above a certain height (different for different ion species), and the ionic fractions `freeze-in' as they do in the solar wind. Therefore, to predict emissions inside the CS, it is important to take this non-equilibrium condition into account, and it is not necessarily valid to calculate the emission based on ionization equilibrium (i.e. only based on the electron temperature). Similarly, the electron temperature derived from line ratios assuming ionization equilibrium may not be the actual electron temperature. 

Figure 6 compares $y_q(r)$ for the three cases at DR height of 1.2 \Ros for a few ions. 
We can see that there are significant differences between models of different coronal profiles that would result in different emission intensities at various heights, as well as different frozen-in charge states. Note that even though models N1T1 and N1T2 have almost the same $n_e$, $T_e$, and $v_{out}$ (Fig.4), the different ambient coronal temperatures result in different initial charge states at the location of the SMS crossing, thus different evolution of the ionic fractions as the ions flow along the CS. Therefore the observations can, in principle, be used to constrain these input model parameters. Note that many ions seem to start freezing-in at higher heights than what are usually expected for the normal solar wind (e.g. B\"urgi \& Geiss 1986, Ko et all. 1997), probably due to much higher density inside the CS than in the solar wind. 

\subsection{Model 2: The Streamline Model}

Another model we explored assumes that the plasma that just comes inside the CS through SMS forms its own `streamlines' and does not mix with the material that comes in at other SMS crossings. The properties inside the SMS at a given location would then be an average quantity from these streamlines weighted by the volume each of these streamlines occupies at that location (see below). While the long collisional mean free path suggests the idea of mixing as in Model 1, it is not clear if the actual magnetic field configuration inside the SMS is the same as that in the Petchek's model. Also, MHD turbulence may inhibit the mixing of the plasma and thermal equilibration (turbulent speeds of $\sim$60 km/s have been found in the post-CME CSs by Bemporad (2008)). Therefore we also want to examine how such a `Streamline model', which can be regarded as an extreme opposite case of Model 1, compares with the observations. 

We use the same three coronal models (i.e. N1T1, N1T2, N2T2), and the same compression/heating/acceleration at the SMS crossing (Eqs.(1), (2), (4)). As the plasma flows in each streamline after the SMS crossing (cp. Fig.1b), the electron density decreases due to spherical expansion:
\begin{equation}
n_e(r)=n_{e,SMS}{r_{SMS}^2\over r^2}
\end{equation}
where $n_{e,SMS}$ (same as $n_2$ in Eq.(1)) is the electron density at the SMS crossing at height $r_{SMS}$. The evolution of the electron temperature is governed by adiabatic cooling and radiative cooling:
\begin{equation}
{\partial T_e(r)\over \partial r} = -{4\over 3}{T_e\over r} - {{2n_en_H\Lambda} \over {3(1.92n_e k u(r))}}.
\end{equation}
where $\Lambda$ is the radiative cooling rate. Because the ions are most likely not in ionization equilibrium along the flow (cp.Eq.(10) and Fig.5), we use the non-equilibrium ionic fractions (Eq.(10)) to calculate the radiative cooling rate. The radiative cooling includes bound-bound, bound-free and free-free processes using routines provided by CHIANTI atomic database version 6.0 (Dere et al. 1997, 2009) but with ionic fractions calculated here replacing those in ionization equilibrium. We find that the radiative cooling is negligible compared to adiabatic cooling in almost all cases except at the few lowest heights where the density is high. Therefore it can be neglected in general. The flow speed in each streamline is equal to the coronal Alfv\'en speed at the SMS crossing (Eq.(4)). The evolution of the ionic fractions in each streamline thus can be calculated according to Eq.(10). Note that we neglect heat conduction in Eq.(12) assuming that MHD turbulence inside the CS suppresses the thermal conduction. Long current sheets are expected to
undergo tearing instability, which can create a large number of
plasmoids at different scales (e.g. Shibata \& Tanuma 2001, B\'arta et al. 2010, and references
therein). Due to the poloidal field of the plasmoids, they are thermally `insulated' in such medium that is filled with large
number of stochastically moving plasmoids, thus reducing the thermal conduction. There will be thermal conduction at some level in spite of the effects of turbulence, but we have chosen the extreme case to investigate the range of possible results.

Once the electron density, temperature and ion outflow profiles in each streamline are known, the ionic fraction profiles in the streamline can be calculated according to Eq.(10). The properties inside the CS at a certain location, in the context of this model, are therefore average quantities taking into account all `streamlines' that form between the DR and that given location, assuming that the streamline structure inside the CS can not be resolved. Under spherical geometry, the volume at height $r$ occupied by a streamline that entered the CS (i.e. crossing the SMS) at height $r_{SMS}$ is proportional to $r/r_{SMS}$. The average quantity of a parameter $X$ observed at height $r$ is thus
\begin{equation}
X(r)={{\Sigma_{r_{SMS}=r_{DR}}^{r_{SMS}=r}  X_{SMS}(r) F(r,r_{SMS})}\over{\Sigma_{r_{SMS}=r_{DR}}^{r_{SMS}=r} F(r,r_{SMS})}}.
\end{equation}
where $r_{SMS}$ and $r_{DR}$ are the height of the SMS crossing and diffusion region, respectively. $X_{SMS}$ is the physical quantity at height $r$ in the streamline that enters the SMS at $r_{SMS}$, and $F(r,r_{SMS})$ is the `contribution factor' from the streamline forming at $r_{SMS}$. $F(r,r_{SMS})$ is different for different physical parameters. For the electron density, $F(r,r_{SMS})\equiv r/r_{SMS}$. For the ionic fraction $y_q$, $F(r,r_{SMS})\equiv rn_e(r)/r_{SMS}$ (ion density is proportional to $y_q*n_e$).

Figure 7 shows an example of how different streamlines contribute to an ionic fraction observed at height $r=1.7$ \Ro~for the case of $r_{DR}=1.2$ \Ro. All 3 coronal models are plotted for comparison. Also plotted are those for the electron density. We choose those ions that emit high-temperature lines observable by UVCS in the CS at this height (i.e. Fe$^{+17}$([Fe XVIII] $\lambda$974), Si$^{+11}$(Si XII $\lambda$499), Ca$^{+13}$([Ca XIV] $\lambda$943) (e.g. Ciaravella et al. 2002, Ko et al. 2003, Bemporad et al. 2006). 
One can see that the streamlines (i.e. $r_{SMS}$'s) that make a major contribution to the average ionic fraction are different for different ions and different coronal models, but most of the contribution is from SMS crossings at low heights. Similar examination for an observed height of 1 AU indicates that the major contribution for most abundant ions at 1 AU is from streamlines that enter the SMS below 2 \Ro. For the electron density, it is the same for all 3 models because $n_e(r)$ evolves in the same way with $r$ (Eq.(11)).

Figure 8 plots the average ionic fractions (Eq.(13)) for selected ions as a function of height for the 3 coronal models. Comparing with the Fully-mixed model, we find that the results of the two models are similar within an order of magnitude. Figure 9 plots the average electron density compared with the data (cp.Fig.4). Fig.9 shows that, similar to the Fully-mixed model, models N1T1 and N1T2 with reasonable DR heights predict electron densities that agree with the observations, while model N2T2 seems to underestimate the electron density. 

\section{Predicted Observables from Model Calculations}

The main purpose of this work is to test post-CME CS models and see if model predictions agree with observations. Figs.4 and 9 show that the predicted electron density profiles within the CS overall agree with the data for both model approaches (see also Fig.7 of Vr\v{s}nak et al. 2009). The two models predict similar electron densities at low heights, but the electron density profile falls faster with height for the Fully-mixed model at $r \gsim 4$ \Ro. Therefore CS observations over an extended range of heights may be able to distinguish such different model approaches and indicate which one represents better the actual conditions inside the CS. It is also possible, in the context of either model, to constrain the coronal parameters outside of the CS. 

\subsection{UV Line Emission}

Another predicted quantity is the line emission (see Sec.1). In coronal conditions, the electron density is low enough that most emission lines are produced by electron collisional excitation followed by spontaneous emission. For a plasma with electron temperature $T_e$, the line emission is thus:
\begin{equation}
I_{line}= {1\over 4\pi}{n_{el}\over n_H} \int{G(T_e) dEM(T_e)}
\phantom{-} {\rm photon ~cm^{-2}s^{-1}sr^{-1}}
\end{equation}
where $n_{el}/n_H$ is the elemental abundance relative to hydrogen (i.e. absolute abundance, Grevesse et al. 2007). $G(T_e)$ is the contribution function which is defined as
\begin{equation}
G(T_e)={n_{ion}\over n_{el}} B_{line} q_{line}(T_e)
\end{equation}
where $n_{ion}/n_{el}$ is the ionic fraction calculated from Eq.(10).
$B_{line}$ is the branching ratio for the line transition, and $q_{line}(T_e)$ is the electron excitation rate
which is a function of only the electron temperature in the low-density limit.
$dEM(T_e) = d(n_e n_H L)$ is the emission measure (in cm$^{-5}$) at a given electron temperature with line-of-sight (LOS) depth $L$, and $n_H=n_e/1.2$ for fully ionized plasma.
In our models, the ionic fractions are not in ionization equilibrium (e.g. see Fig.5). Therefore we calculate $G$ from the CHIANTI atomic database version 6.0 (Dere et al. 1997, 2009) but with the ionic fractions calculated here replacing those in ionization equilibrium (e.g. Figs.6 and 8).

For the Fully-mixed model, the line emission can be calculated directly from Eq.(14) with the electron density, temperature and ionic fractions calculated in the model. For the Streamline model, one needs to calculate the average quantity as in Eq.(13) with $X_{SMS}(r) \equiv I_{line,SMS}(r)$ and  $F(r,r_{SMS})\equiv r/r_{SMS}$. There are two assumptions needed. One is the elemental abundance, and the other is the LOS depth. The First Ionization Potential (FIP) effect is commonly observed in coronal plasmas. The FIP bias, defined as the abundance ratio of a low-FIP element (FIP $<$ 10 eV) to a high-FIP element (FIP $>$ 10 eV) relative to its photospheric ratio, is generally in the range of 3-4 in active regions and streamers (e.g. see review by Raymond et al. 2001 and references within). Since fast CMEs with such WL ray/CS structure almost always occur from active regions, we assume that the abundance for all low-FIP elements is $3\times$ its photospheric value, and the abundance for all high-FIP elements is equal to its photospheric value (Grevesse et al. 2007). For the LOS depth, we adopt the value of 0.05 \Ro~at 1.1 \Ro~so the LOS depth at a given height $r$ is $0.05 \times r/1.1$ \Ro~under spherical expansion. For individual CME events, this LOS depth can be estimated from the length and inclination of the neutral line at the eruption site (Ciaravella \& Raymond 2008, Vr\v{s}nak et al. 2009, Lin et al. 2009). In any case, different assumptions for the FIP bias and LOS depth can be easily taken into account since the calculated line fluxes are linearly proportional to these two quantities.

Figure 10 plots the predicted fluxes for Si XII $\lambda$499, [Ca XIV] $\lambda$943 and [Fe XVIII] $\lambda$974 versus heliocentric heights for the Fully-mixed model with 3 coronal models and DR height at 1.2 \Ro. As previously mentioned, these lines are chosen because they were commonly observed by UVCS in post-CME CS events. UVCS data for 4 events are also plotted for comparison. Fig.11 plots the same but for the Streamline model. The two models predict similar line fluxes at low heights ($\lsim$ 2.0 \Ro). The difference is at higher heights where the line fluxes in the Fully-mixed model decrease more rapidly with height. At 5 \Ro, these lines are fainter by an order of magnitude in the Fully-mixed model than the Streamline model. At heights of the available UVCS observations ($\le$ 1.7 \Ro), the model predictions agree with the observations in an overall sense. The Mar.23, 1998 event seems to be more consistent with the N1T1/N1T2 models, and the Jan.8, 2002 event seems to be more consistent with the N2T2 model. This would indicate different ambient coronal conditions for different events. For the Nov.4, 2003 event, however, Si XII and Fe XVIII lines are consistent with different coronal models. One possible explanation is that none of the coronal models adopted here represents the actual coronal condition for this  event. A higher temperature in CS (e.g. from smaller coronal $\beta$, Eq.(2)) would produce less Si XII emission relative to the Fe XVIII emission. For Ca XIV, all models underestimate the line emission up to a factor of 5. This line is blended with [Si VIII] $\lambda$944 but the absence of low-temperature emission along the slit where the Fe XVIII emission exists rules out line blending as a possible cause. The discrepancy may lie in the atomic data and the assumed coronal abundance.   

Figure 12 shows an example of how different streamlines contribute to the line fluxes observed at height $r=1.7$ \Ro~for the case of $r_{DR}=1.2$ \Ro. All 3 coronal models are plotted for comparison. Depending on the individual line, the dominant emission comes from `streamlines' of a limited range of the SMS crossings. There is little difference among the 3 coronal models. This is in contrast with the ionic fractions (Fig.7), probably because of a combination of the effects of the electron density and temperature on the line emission. 

One interesting aspect to examine is the change of line emission with time. As the reconnection continues, the location of the DR will move up toward higher height with speed approximately equal to the inflow speed of the reconnection (Lin \& Forbes 2000). Since the formation of the SMS is much faster (in Alfv\'en speed) than the movement of the DR, the SMS developed from a given DR height can be regarded as a steady state structure. The properties within the CS will then change with the movement of the DR. The time evolution of these properties may provide information about the DR movement, therefore the inflow speed/reconnection rate. Figures 13 and 14 plot the 3 line fluxes and the electron density versus DR height as seen at 3 locations (1.5, 1.6, 1.7 \Ro), along with the UVCS data, for the Fully-mixed model and Streamline model, respectively. The time sequence of line fluxes and electron density in the Nov.4, 2003 event (purple symbols, Ciaravella et al. 2008, observed at 1.66 \Ro) indicates that during 2.5 hours, the Fe XVIII fluxes decreased by a factor of 0.63, Si XII fluxes increased by a factor of 1.5, and the electron density decreased by a factor of 0.7. If we use model N1T2 as a proxy (note that the direction of change in these 3 quantities is the same as the data, even though the magnitude does not all fit), this roughly corresponds to an upward motion of around 0.15 \Ro~which in turn corresponds to an inflow speed of $\sim$10 km/s and $M_A$ $\sim$0.01 (based on Alfv\'en speed of $\sim O(10^3)$ km/s). This is consistent with both theoretical (although on the low side, e.g. Lin \& Forbes 2000) and observational (e.g. Yokoyama et al. 2002, Ko et al. 2003, Lin et al. 2005) findings.  


\subsection{Simulated Broad-Band X-Ray Emission}

Besides particular lines of interest shown above, our model of the electron density, temperature and ionic fractions for the 13 most abundant elements enables us to calculate emission spectra across a continuous wavelength range. One such application is to use the emission code of Raymond \& Smith (1977; updated
in Cox \& Raymond 1985) which calculates the line and continuum emission spectrum. This predicted spectrum can then be compared with observations from broad-band instruments such as Hinode/XRT and SDO/AIA, after folding with the instrument's wavelength response function, for specific CS events these instruments might observe.

To demonstrate this, we use {\it the non-equilibrium ionization states} computed with both models
along with the elemental abundances described above to compute the emission over the
range 2.5-4998 eV (2.48-4960 \AA ). The Raymond \& Smith model includes fewer emission
lines than the APEC code (Smith et al. 2001), but for low spectral resolution data
the results are very similar. We then fold the emission spectrum with the wavelength response function of the XRT instrument on Hinode to obtain the predicted count rate of the CS if it were to be seen by XRT. The wavelength response function is calculated using the standard SolarSoft routines with contamination thickness for Apr.9, 2008, 14 UT. We chose this particular instrument and date because a very interesting CME/CS event was observed by SOHO and Hinode on this day that has been analyzed in great detail (Savage et al. 2010, Landi et al. 2010). 

Figures 15 plots the predicted count rate (DN/sec/pixel) versus height for the Fully-mixed model as seen by Hinode/XRT thin Al-Poly filter for the 3 coronal models and 3 DR heights. Figure 16 plots the same for the Streamline model.
The LOS depth is assumed to be the thickness of the CS, $5\times 10^3$ km, as determined by Savage et al. (2010) for the Apr.9, 2008 event. 
We can see that the count rate can differ by an order of magnitude among coronal models and is mainly dictated by the ambient coronal density, not the electron temperature. Figure 17 shows the CS observed by Hinode/XRT on Apr.9, 2008. Plotted in the upper panels are the measured count rates along the solar Y-position at 4 solar X-positions that cross the CS (marked in the lower panels). The location of the DR is at $\sim$1.25 \Ro~based on the downflows/upflows measured by Savage et al. (2010). This is around the second outermost X-position marked in Fig.17 (inner solid line). The `bumps' between Y pixel of 100-300 are where the CS is located. Plotted in red on the two lowest curves (solid lines) are the Y-pix ranges chosen to estimate the mean and standard deviation of the count rate within the CS. These numbers are shown as blue data points on Figs.15 and 16. The heliocentric heights of these data points were estimated based on that the flare/CME source was $\sim 23^\circ$ behind the limb at 14 UT on Apr.9, 2008 (Savage et al. 2010). Comparing the predicted and observed count rates at these X-positions indicates that our model calculations are in the agreeable range with the data for the N1T1 and N1T2 models, but not for the N2T2 model. Note that our predicted count rates, bracketing the data between DR height of 1.2 and 1.3 \Ro, are also consistent with the position of the X point ($\sim$ 1.25 \Ro) as found out by Savage et al. (2010). The agreement implies that the ambient coronal density profile is close to the `N1' profile, at least at around 1.3 \Ro. It should be possible to compare the `N1' profile with the electron density profile derived from the white light polarization brightness (pB) measurement at the time before and during the event and see if they agree, but we do not make such attempt in this paper.  

\subsection{Solar Wind Charge States at 1 AU}
 
Many of the magnetic clouds (MCs) seen as interplanetary coronal mass ejections (ICMEs) bear high ion charge states usually represented in high O$^{+7}$/O$^{+6}$ ratios ($>$ 0.2) and high-ionization charge states of Fe, e.g. Fe$^{+16}$ (Lepri \& Zurbuchen 2004). Since the ion charge states usually freeze-in near the Sun, this implies electron heating in the CME material in the early stage of the eruption process (e.g. Akmal et al. 2001, Rakowski et al. 2007, Lee et al. 2009). Our model of the post-CME CS also predicts higher charge states due to higher electron temperature than the `normal' corona from the electron heating crossing the SMS, as shown in our results. Plasma ejected upwards in the current sheet enters the flux rope formed by reconnection below the CS core.  If there is a pre-existing flux rope, reconnection forms a flux rope around it having a comparable amount of magnetic flux, and therefore
volume (Lin et al. 2004, M\"{o}stl et al. 2008), so that much of the plasma in a magnetic cloud has passed through the current sheet. However, there is no report so far of definite signatures of post-CME CS measured in-situ at 1 AU. Note that this `CS' is different from the `in-situ reconnection exhaust' reported by Gosling et al. (2005), but the model work presented here can in principle be applied to this case with very different in-situ ambient conditions outside the SMS.

Figures 18 and 19 show the predicted frozen-in O$^{+7}$/O$^{+6}$ ratios and the average Fe charge at 1 AU for the Fully-mixed model and Streamline model, respectively. We can see that there are notable differences between the two model approaches, as well as the 3 coronal models. 
In general, the Streamline model predicts higher charge states at 1 AU (or $>$ 20 \Ro~according to Fig.6 and 8) than the Fully-mixed model. The ionization charge state is higher for model N2T2 than the other two coronal models due to higher CS temperature (e.g. Fig.4), as opposed to that the emission is lower for model N2T2 due to lower density. We note that the range of these predicted quantities of O$^{+7}$/O$^{+6}$ and average Fe charge is common in the ICME solar wind data (O$^{+7}$/O$^{+6} > 0.2$, average Fe charge $> 12$. See, e.g. Lepri \& Zurbuchen 2004, Wimmer-Schweingruber et al. 2006, Zurbuchen \& Richardson 2006). If some of the ICMEs with high O$^{+7}$/O$^{+6}$ and Fe charge states are associated with such post-CME CS structure, It would imply that the diffusion region remains low in the corona. It remains to be seen if the models presented here can be tested by observations of charge states within the to-be-identified post-CME CS structure in the solar wind.

\section{Discussion}

In order to understand post-CME WL rays and current sheets, it is important to have
quantitative comparisons between theory and observations.  The models presented here are
a first step in that direction.  Several crucial observations exist, including the
Fe XXIV spines seen above some post-flare loops in TRACE images (Innes et al. 2003a),
the high velocity Fe XIX and Fe XXI lines seen by SUMER in several events
(Innes et al. 2003b; Wang et al. 2007), the UVCS observations of highly ionized
gas (Ciaravella et al. 2002, Ko et al. 2003, Bemporad et al. 2006, Ciaravella \& Raymond 2008), evidence of reconnection downflows/upflows observed by Hinode/XRT (Savage et al. 2010), and {\it in situ} measurements of high ionization states in ICMEs (e.g. Lepri \& Zurbuchen 2004).

We have taken a simple model of the exhaust region of Petschek reconnection and
computed the time-dependent ionization state in order to predict UV emission line
intensities and the count rates in X-ray bands.
This model is consistent with observations, thus it is a viable explanation for the observed `post-CME current sheet'. This does not exclude other possible models but any model should be able to predict properties that can produce observables that are consistent with the observations. The Vr\v{s}nak et al. (2009) model, in our opinion, is the most viable model so far that adequately treats plasma properties of the CS {\it above} the DR for which we can calculate physical parameters and compare with the observations. Given the model assumptions, the good agreement with the observations in both density and UV line/X-ray emission is more than just fortuitous.  In the context of this model, the width of the WL ray increases with height which is also consistent with the data (see Vr\v{s}nak et al. 2009) . This indicates that this model can be the right interpretation of the observed WL ray/CS. In principle, our calculations can be applied to any post-CME CS models that provide information about the electron density/temperature and ion flow speed, such as in the MHD models (e.g. Riley et al. 2008 by using proton parameters). Note that the main purpose of this paper is to show how this particular post-CME CS model concept agrees with the observations but we do not attempt to make fine adjustments of the model parameters to fit data of any one particular CS event. It is also possible to apply such calculations to model the downflow region below the DR (e.g. Yokoyama \& Shibata 1998, 2001) but we do not make such attempt in this paper. 

Both CS models (`Fully-mixed' and `Streamline') are found to give similar UV and X-ray emission at heights below $\sim$2 \Ros but it drops more quickly with height in the Fully-mixed model. Thus it seems that CS observations higher than 2 \Ro would be able to differentiate the two models. However, current instruments do not have the sensitivity yet to detect such high temperature emission much beyond what is presented here. Figs.4 and 9 indicate that electron densities derived from WL observations beyond 4 \Ros can be used to differentiate the two models. Such data are available although we do not attempt to acquire in this work. The ionic charge state (Figs.18 and 19) is another useful parameter to differentiate these models. However, an unambiguous, positive identification of such post-CME CS in the solar wind is necessary before such comparison can be made.  

The model results depend strongly on the model of the ambient corona. The differences among the three coronal models show that, within the context of the models here, the coronal electron density profile is the most important factor in determining the emission properties inside the CS. In this work, we did not investigate the effect of the ambient coronal magnetic field. But we expect it will also play a crucial role (as dictating the plasma $\beta$) in determining the CS properties. The more accurately the ambient conditions at the time of the CS event are known, the better such observations can pinpoint the reconnection processes and properties in the post-CME CS. And comparison of such models presented here with observations can offer a possible diagnostic tool for inferring the external coronal conditions of the post-CME CS. 

As a final remark, the calculation of plasma emission and time-dependent ionization rely on atomic rates for collisional excitation, ionization and recombination. Aside from the accuracy issue of these atomic rates, the electron velocity distribution function (VDF) is an important factor in these rate calculations. In this work, we assume that the electron VDF is a Maxwellian. The deviation of the VDF from a Maxwellian can affect these rates in different degrees for different ions (e.g. Ko et al. 1996, Dzif\v{c}\'akov\'a \& Mason 2008). One future effort of our calculations will be to investigate possible effects of a non-Mawellian electron VDF. A kinetic approach in magnetic reconnection/CS models that can provide information about the electron VDF would be helpful in nailing down our knowledge of the post-CME CS with observations.

 
\acknowledgments
We thank ISSI (International Space Science Institute, Bern) for the hospitality provided to the members of the team on the Role of Current Sheets in Solar Eruptive Events, where many of the ideas presented in this work have been discussed. We thank M. Guhathakurta for providing the electron density profile measured from the SPARTAN 201-1 Mission, P. Bryans for providing the programs that calculate the ionization and recombination rates, A. Bemporad for providing the UVCS data analysis results for the Nov.26, 2002 CS event, A. Ciaravella for helpful discussions, and the anonymous referee for valuable comments and suggestions. SOHO is a joint mission of the European Space Agency and United
States National Aeronautics and Space Administration. Hinode is a Japanese mission developed and launched by ISAS/JAXA, with NAOJ as domestic partner and NASA and STFC (UK) as international partners. It is operated by these agencies in co-operation with ESA and the NSC (Norway). CHIANTI is a collaborative project involving the NRL (USA), RAL (UK), MSSL (UK), the Universities of Florence (Italy) and Cambridge (UK), and George Mason University (USA). This work is supported by NRL/ONR 6.1 basic research program. This work was partially supported by NASA Grant NNX09AB17G-R and NNX07AL72G to the Smithsonian Astrophysical Observatory.

\clearpage

\begin{figure}
\epsscale{1.0}
\plotone{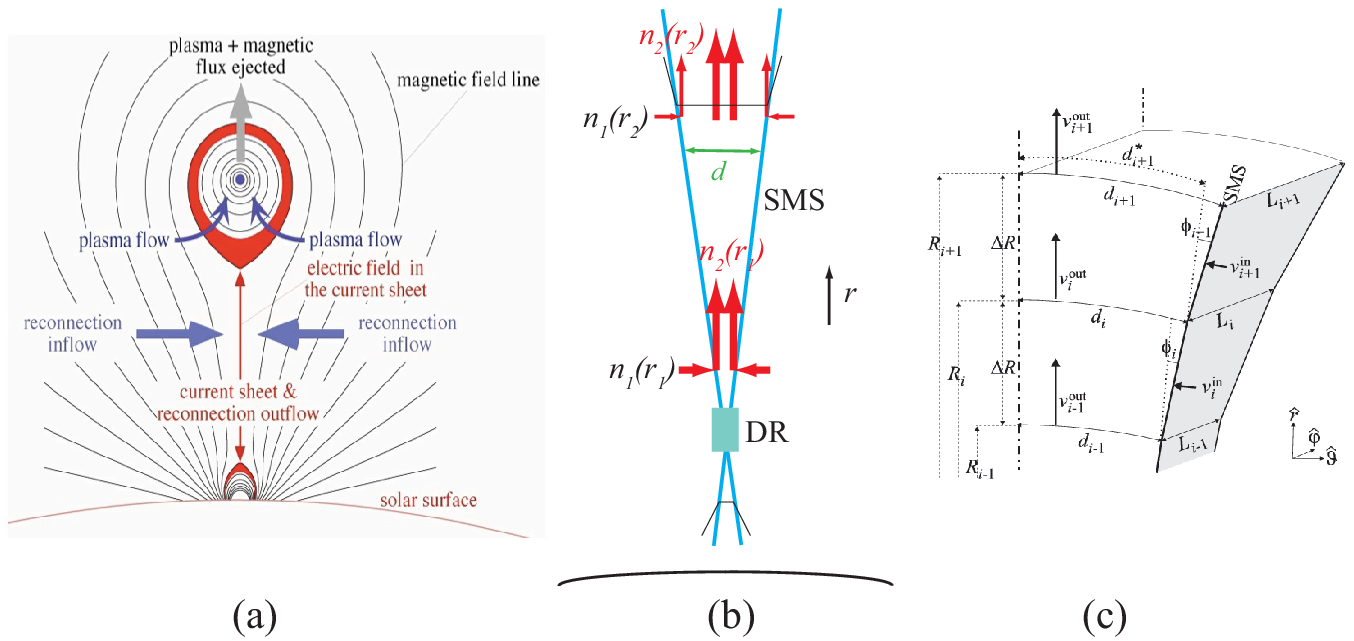}
\caption{(a) Standard flare-CME model depicting a current sheet between the post-flare loops and the ejecting flux rope (e.g. Lin \& Forbes 2000). (b) Sketch of the model by Vr\v{s}nak et al. (2009) that interprets the observed WL ray (CS) as the feature bounded by a pair of slow-mode shocks (SMS) above the diffusion region (DR). (c) Sketch of the `Fully-mixed model' (adopted from Vr\v{s}nak et al. 2009). 
\label{fig1}}
\end{figure}

\begin{figure}
\epsscale{1.0}
\plotone{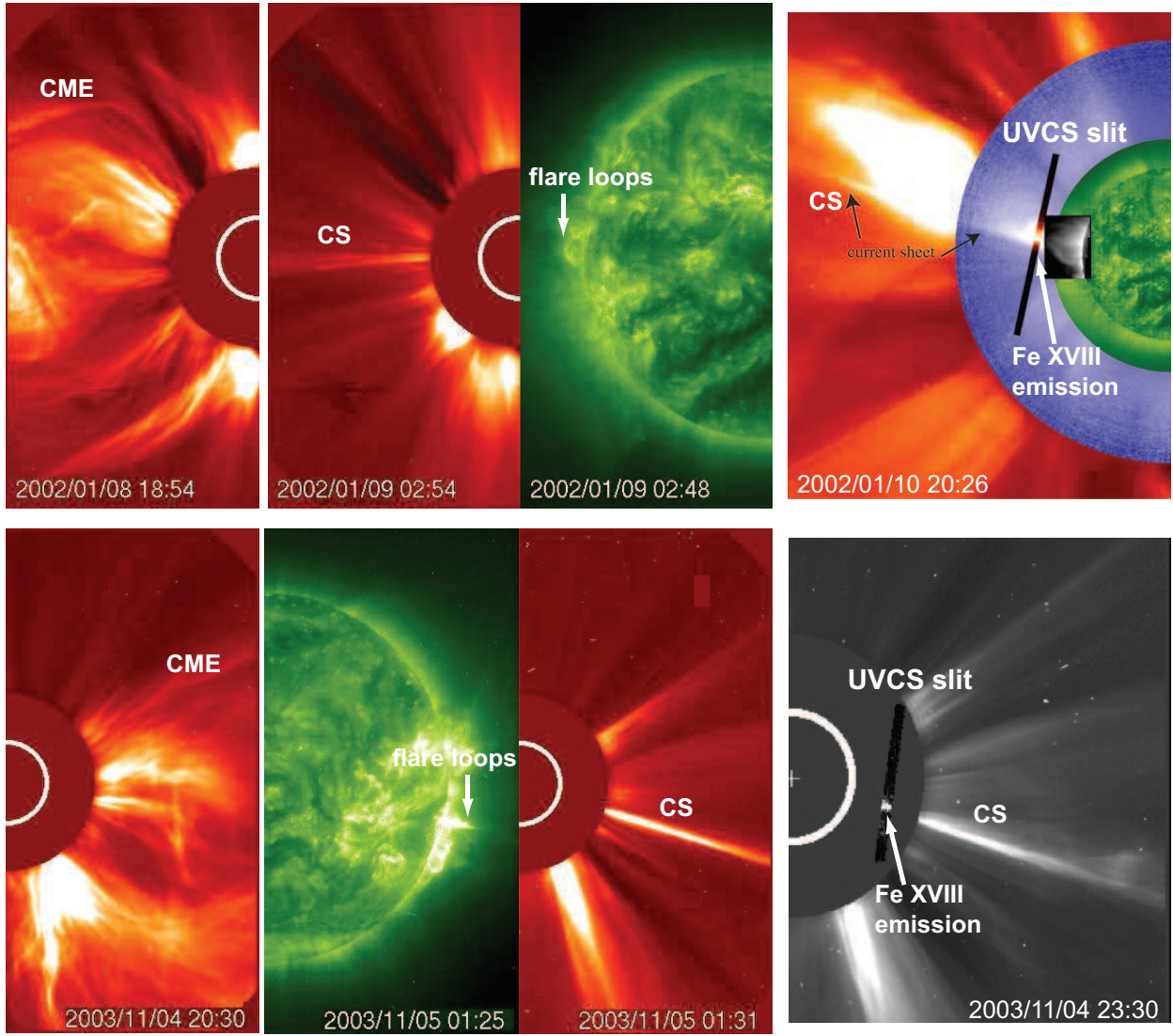}
\caption{Upper panels: the CME/CS event on Jan.08, 2002 (see Ko et al. 2003). Lower panels: the CME/CS event on Nov.04, 2003 (see Ciaravella \& Raymond 2008). Images are from SOHO/LASCO (coronal images in red-orange), SOHO/EIT $\lambda$195 (solar images in green), SOHO/UVCS (slit images marked in the two right panels), SOHO/CDS (coronal loop image in grey in the upper-right panel) and MLSO/MK4 (coronal image in blue in the upper-right panel).
\label{fig2}}
\end{figure}

\begin{figure}
\epsscale{1.0}
\plotone{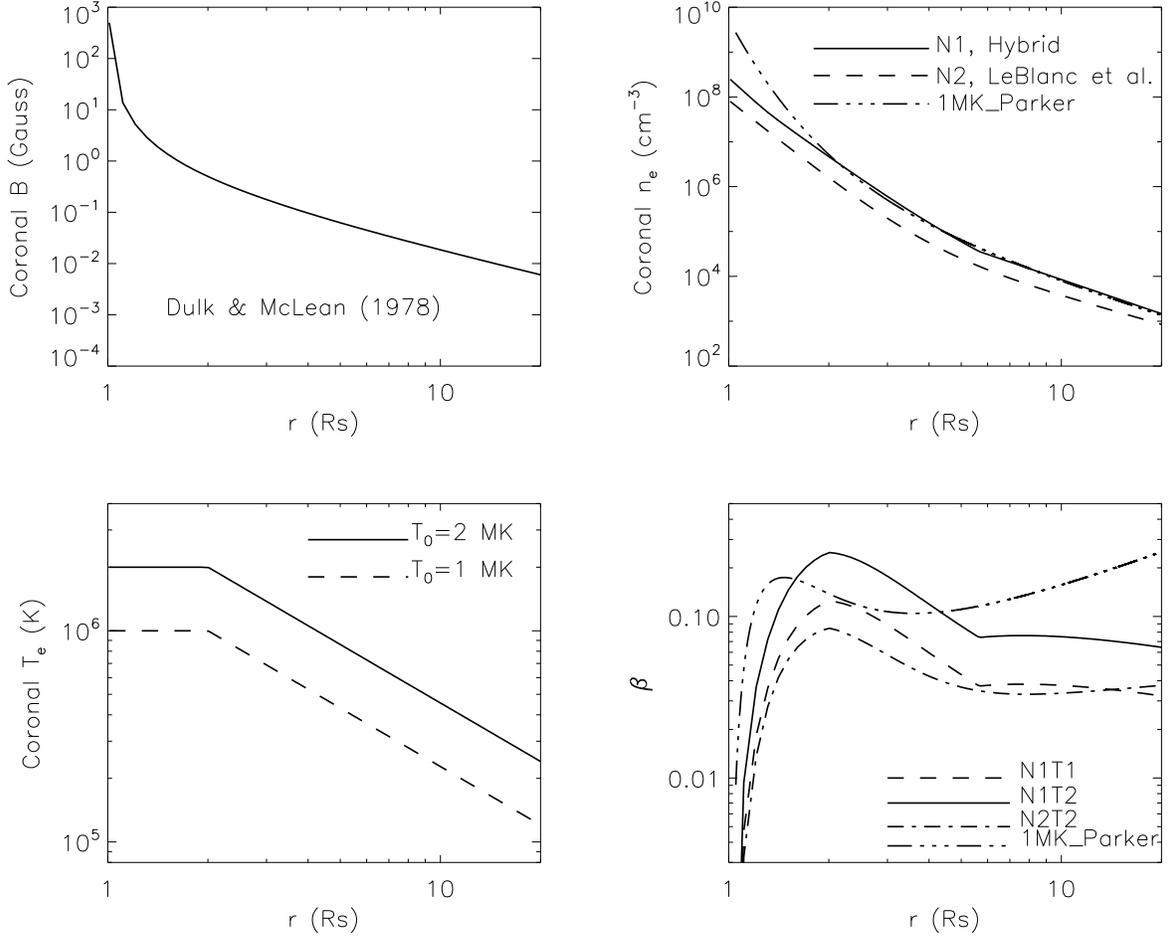}
\caption{Profiles for coronal magnetic field strength (upper left), electron density (upper right), electron temperature (lower left) and plasma beta (lower right) adopted for the study. Also plotted is the electron density and plasma beta profiles for the `1MK\_Parker' model adopted in Vr\v{s}nak et al. (2009). The electron temperature profile for the 1MK\_Parker model is isothermal at $T_e(r)=1\times 10^6$ K.
\label{fig3}}
\end{figure}

\begin{figure}
\epsscale{0.45}
\plotone{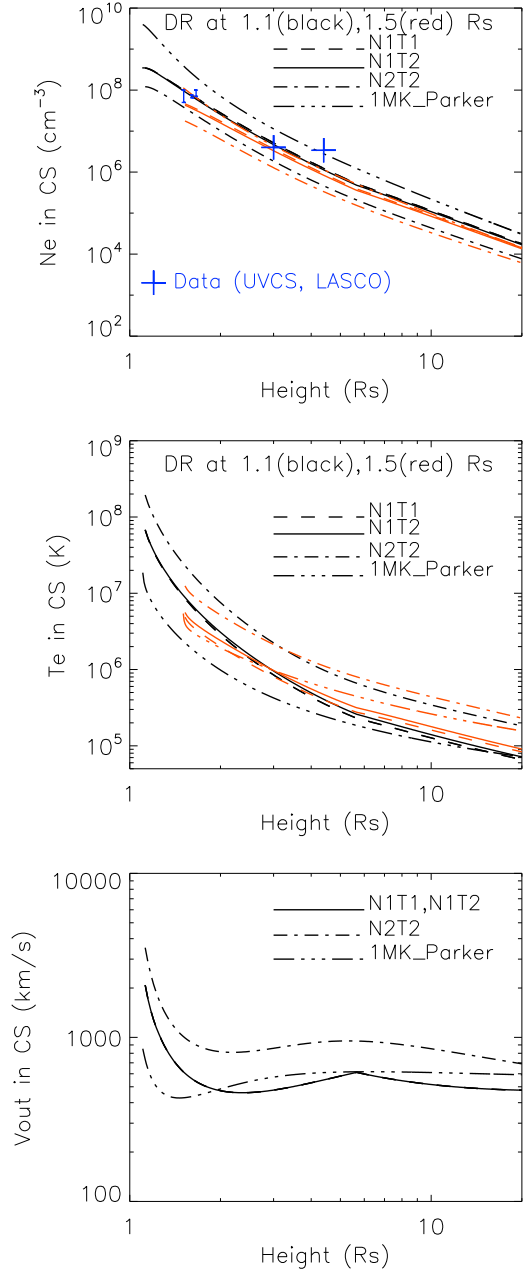}
\caption{$n_e(r)$, $T_e(r)$ and $v_{out}(r)$ for the 3 cases of the coronal $n_e/T_e$ profiles, and for DR at 1.1 (in black) and 1.5 \Ros (in red) for each case. The profiles from Vr\v{s}nak et al. (2009) (`1MK\_Parker', dash-dot-dot-dot) and the electron densities derived from UVCS and LASCO data (blue bars and pluses) are also plotted for comparison. UVCS data are from the Mar.23, 1998 event (Ciaravella et al. 2002) at 1.50 \Ro, Nov.26, 2002 event (Bemporad et al. 2006) at 1.61 \Ro, and the Nov.4, 2003 event (Ciaravella \& Raymond 2008) at 1.66 \Ro. LASCO data are from the Jan.8, 2002 event (Ko et al. 2003) at 3.0 and 4.4 \Ro.  Note that since, with the same coronal field, $v_{out}$ only depends on the ambient density, it is the same for cases N1T1 and N1T2.
\label{fig4}}
\end{figure}

\begin{figure}
\epsscale{0.7}
\plotone{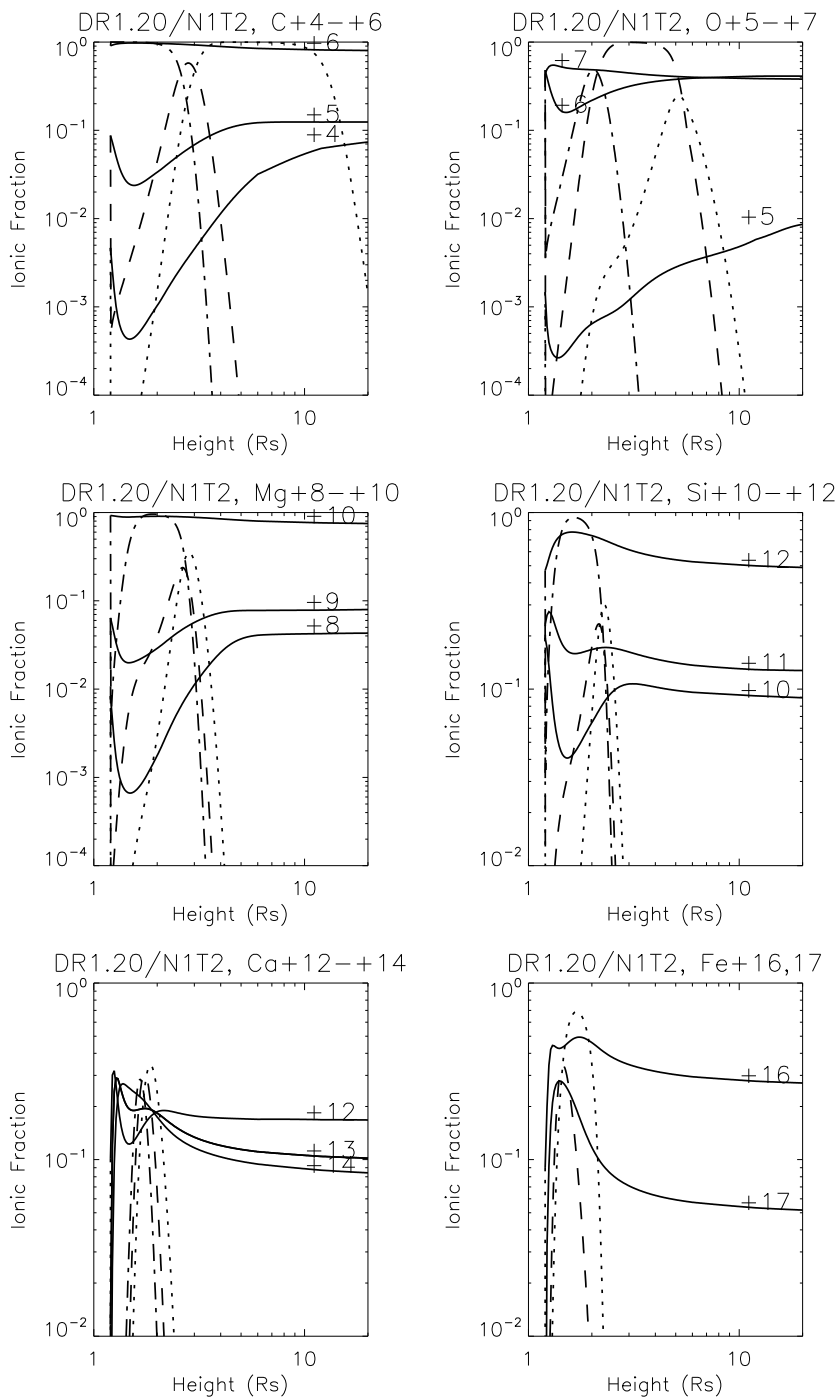}
\caption{Ionic fraction profiles for DR height at 1.2 \Ro~for the Fully-mixed model with coronal model N1T2, compared with those in ionization equilibrium at the local electron temperature in the CS (dotted, dash, dash-dot in the order of ascending charge state).
\label{fig5}}
\end{figure}

\begin{figure}
\epsscale{1.0}
\plotone{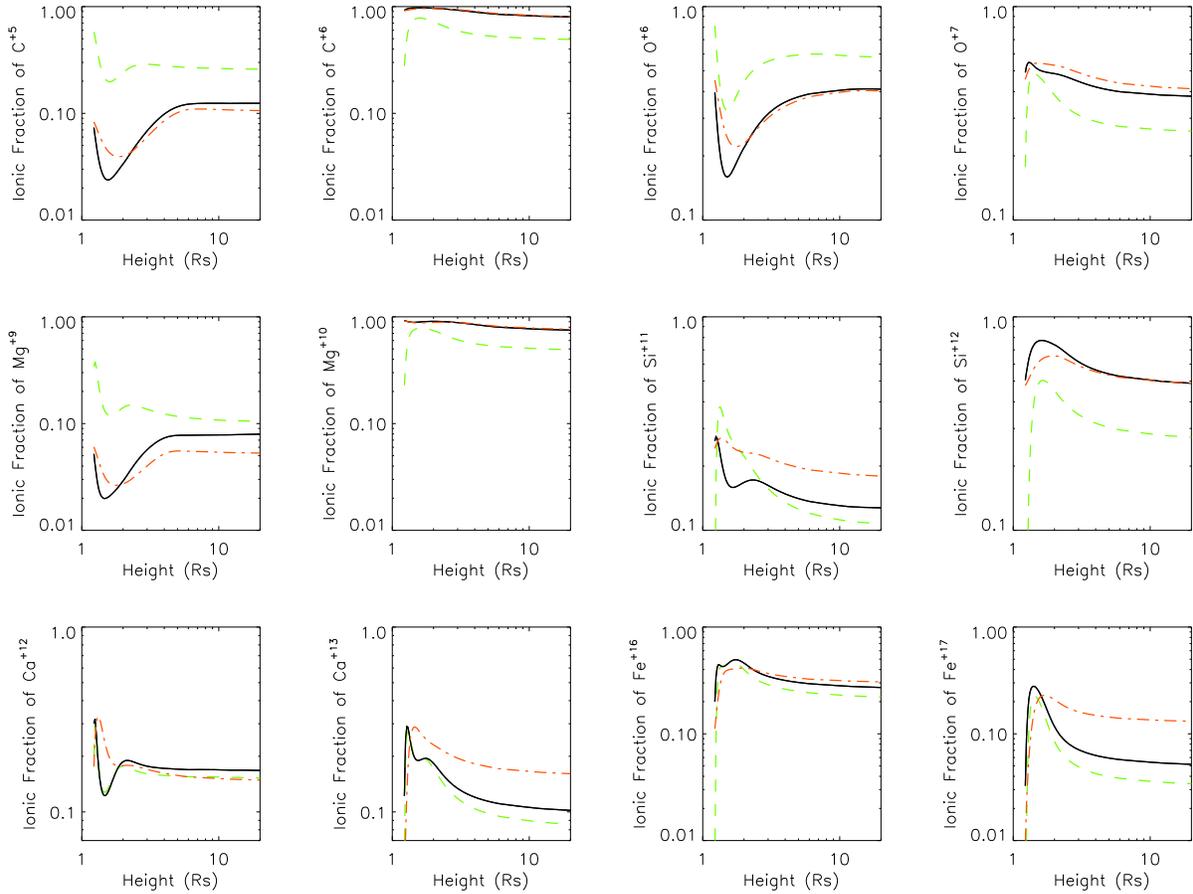}
\caption{Ionic fraction profiles for DR height at 1.2 \Ro~for the Fully-mixed model. Plotted are for 3 coronal models, N1T1 (green-dash), N1T2 (black-solid), and N2T2 (red-dash-dot).
\label{fig6}}
\end{figure}

\begin{figure}
\epsscale{0.7}
\plotone{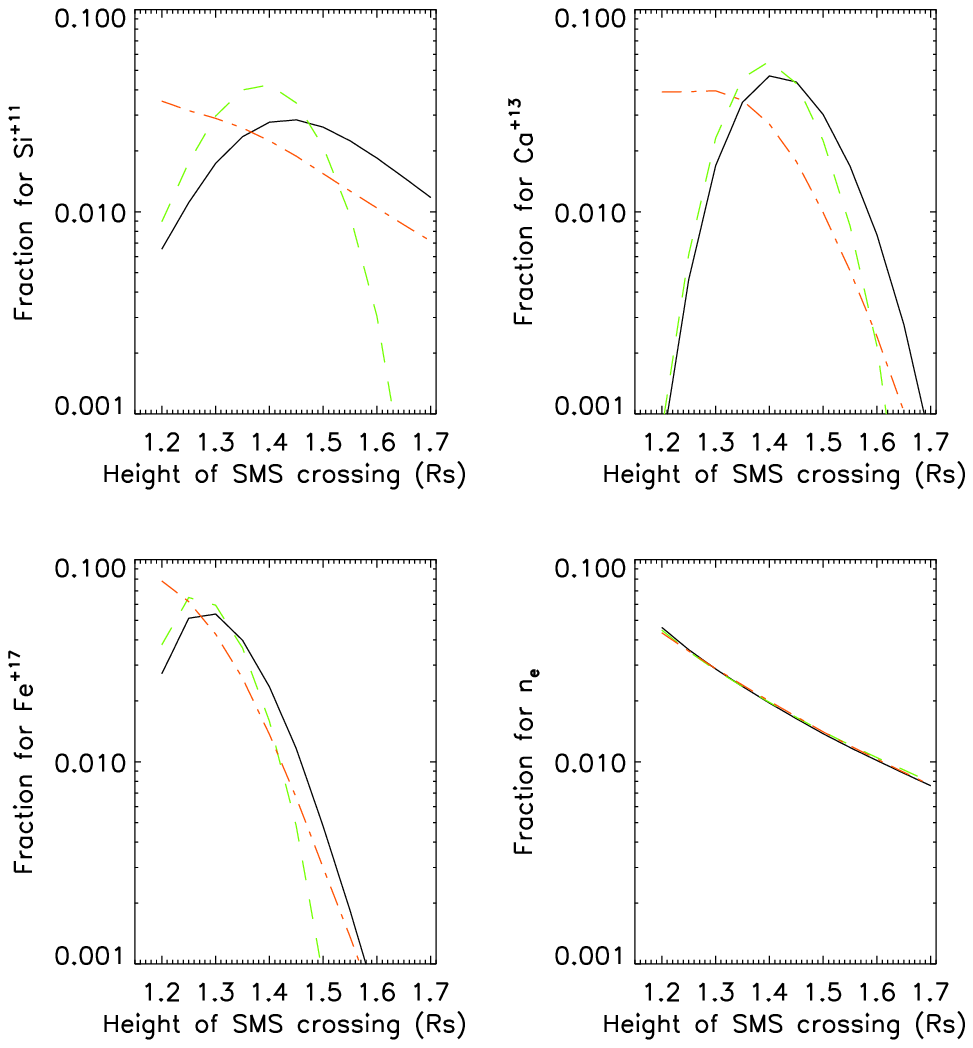}
\caption{Contribution to the total ionic fraction observed at height 1.7 \Ro~from streamlines entering the SMS at heights from 1.2 \Ro~(DR height) to 1.7 \Ro. Plotted are for 3 ions and the electron density for 3 coronal models, N1T1 (green-dash), N1T2 (black-solid), N2T2 (red-dash-dot). The SMS height step is 0.01 \Ro.
\label{fig7}}
\end{figure}

\begin{figure}
\epsscale{1.0}
\plotone{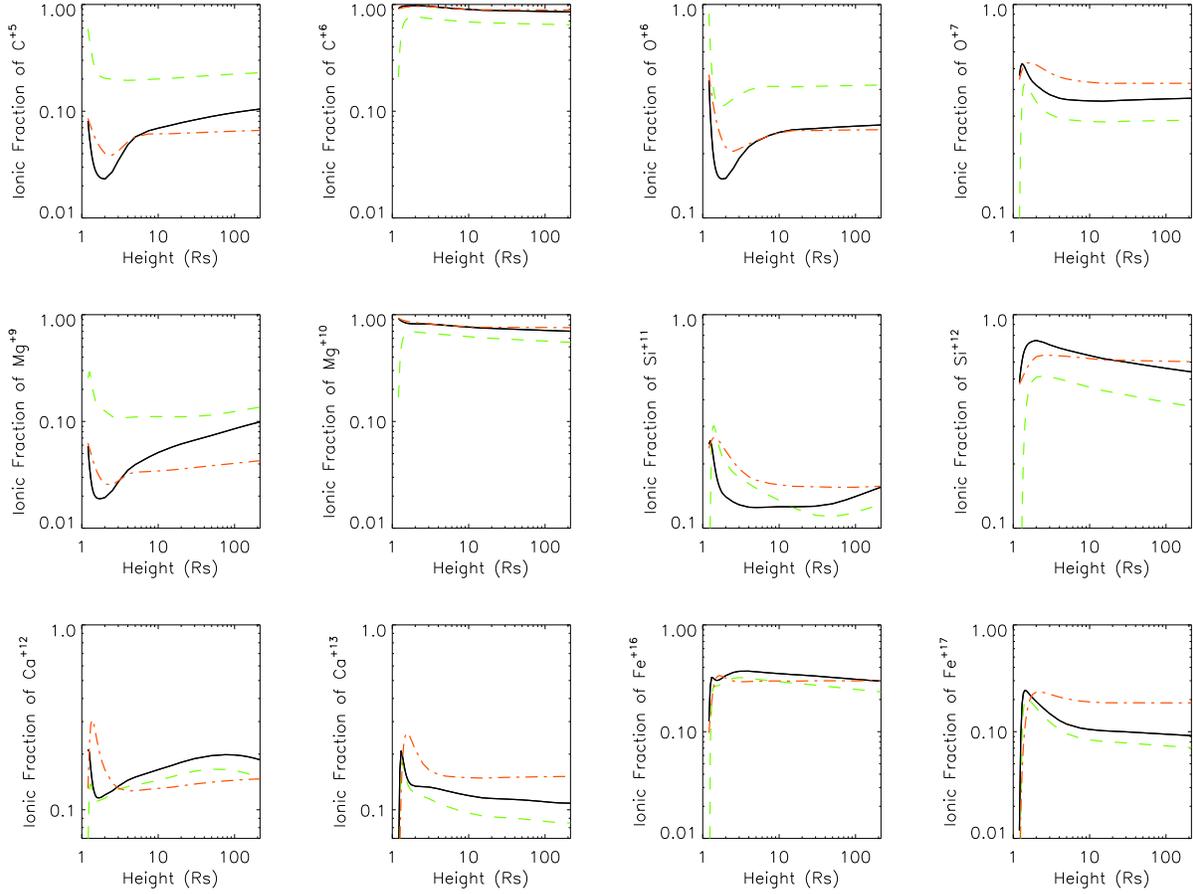}
\caption{Ionic fraction profiles for DR height at 1.2 \Ro~for the Streamline model. Plotted are for 3 coronal models, N1T1 (green-dash), N1T2 (black-solid), N2T2 (red-dash-dot). 
\label{fig8}}
\end{figure}

\begin{figure}
\epsscale{0.7}
\plotone{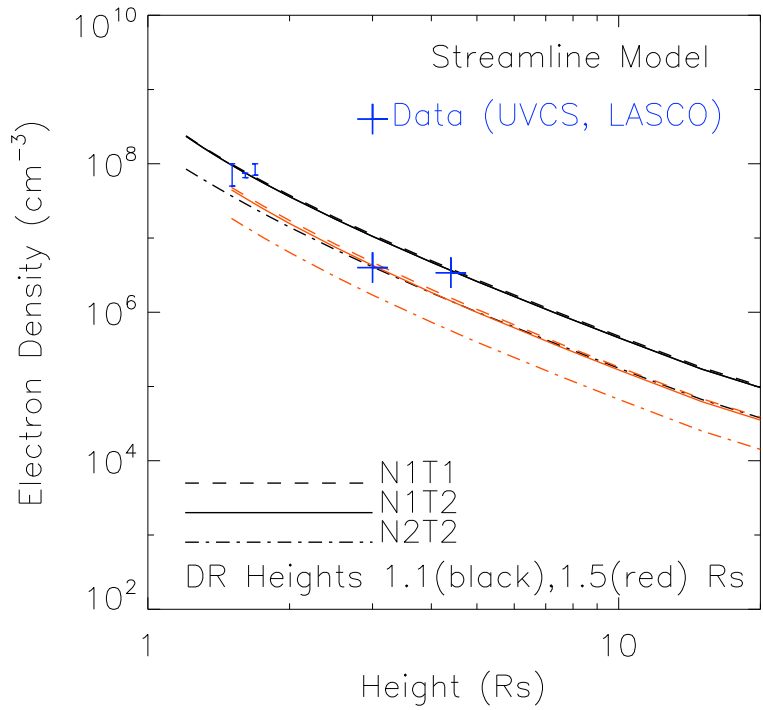}
\caption{Electron density profiles for DR height at 1.1 (black) and 1.5 (red) \Ro~for the Streamline model. Plotted are for 3 coronal models, N1T1 (dash), N1T2 (solid), N2T2 (dash-dot). Also plotted for comparison are $n_e$ derived from UVCS and LASCO data (see caption of Fig.4).
\label{fig9}}
\end{figure}

\clearpage

\begin{figure}
\epsscale{1.0}
\plotone{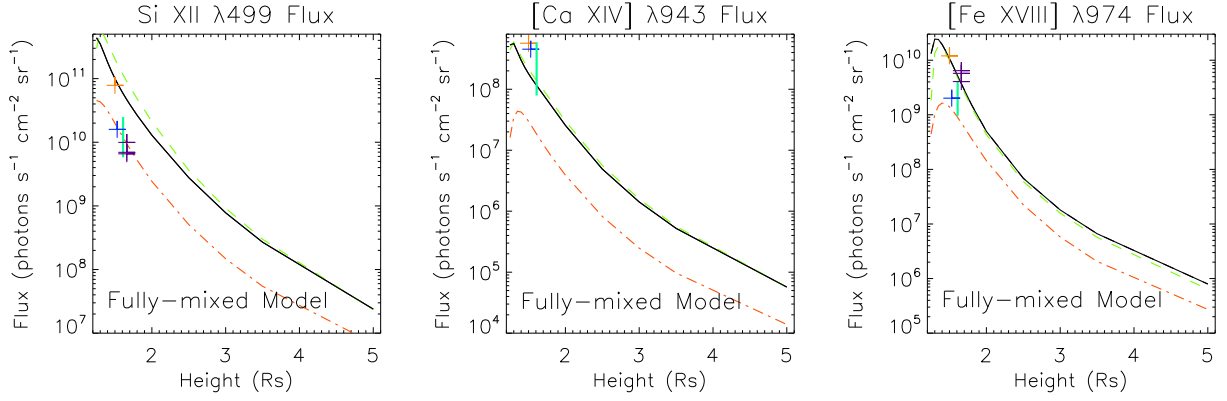}
\caption{For the Fully-mixed model: Three line fluxes versus height for 3 coronal models, N1T1 (green-dash), N1T2 (black-solid), N2T2 (red-dash-dot) for DR height at 1.2 \Ro. Also plotted are the data from UVCS for the CS events on Mar.23, 1998 (orange) at 1.50 \Ro, Jan.8, 2002 (blue) at 1.53 \Ro, Nov.26, 2002 (green, shown is the range from Nov.27, 00:26 UT to Nov.29, 00:20 UT) at 1.61 \Ro, and Nov.4, 2003 (purple, 3 points from Nov.4, 21:06 UT to Nov.5, 00:12 UT) at 1.66 \Ro. The coronal abundance is assumed to be $3\times$ and $1\times$ the photospheric value for low-FIP and high-FIP elements, respectively, and the LOS depth is assumed to be 0.05 \Ro~at 1.1 \Ro. Note that for the Si XII flux, that of the Jan.8, 2002 event has the background corona taken out. Others should be regarded as upper limit.
\label{fig10}}
\end{figure}

\begin{figure}
\epsscale{1.0}
\plotone{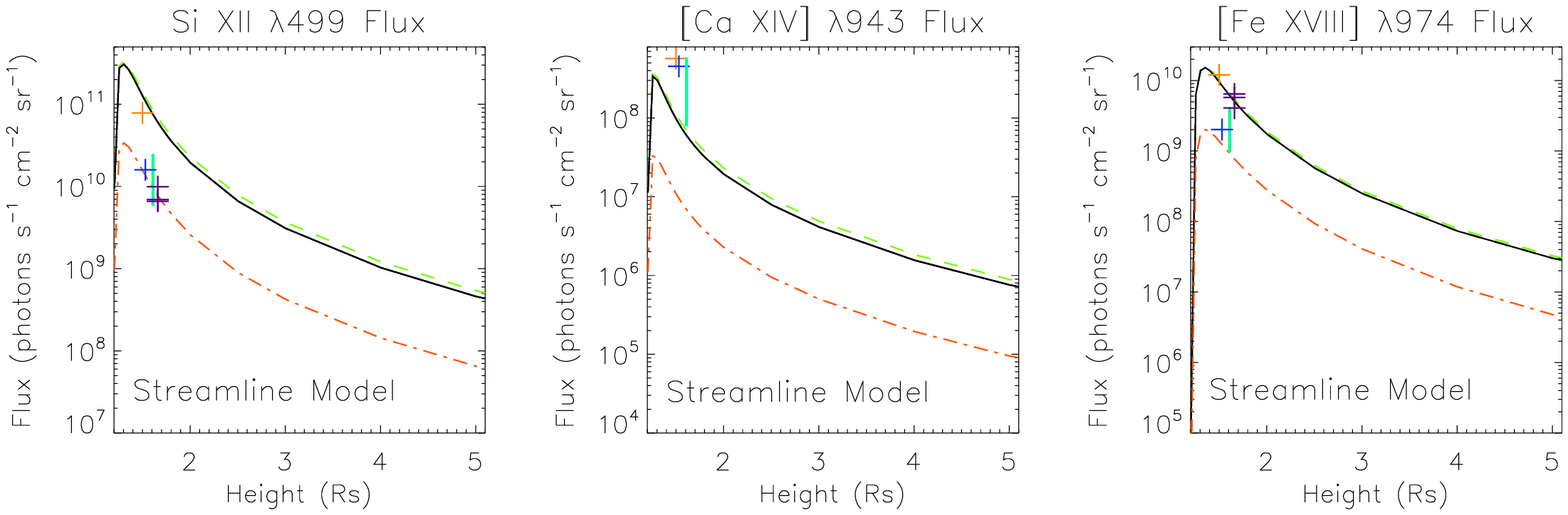}
\caption{Same as Fig.10 but for the Streamline model.
\label{fig11}}
\end{figure}

\begin{figure}
\epsscale{1.0}
\plotone{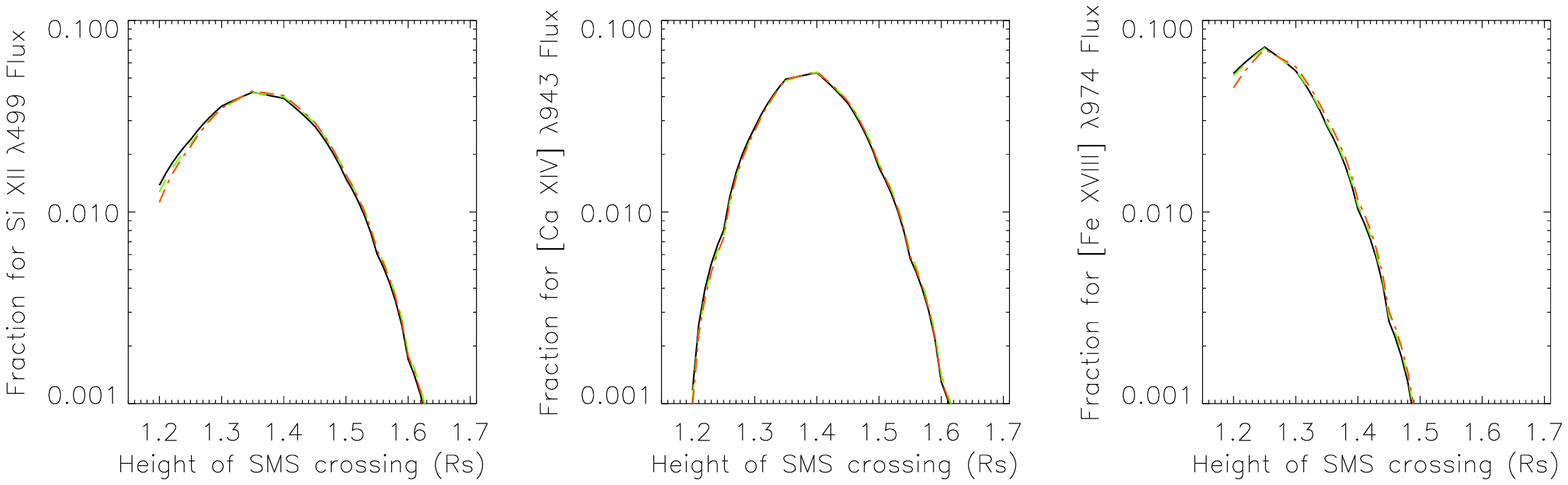}
\caption{Contribution to the total flux observed at height 1.7 \Ro~from streamlines entering the SMS at heights from 1.2 \Ro~(DR height) to 1.7 \Ro. Plotted are for 3 lines for 3 coronal models, N1T1 (green-dash), N1T2 (black-solid), N2T2 (red-dash-dot). The SMS height step is 0.01 \Ro.
\label{fig12}}
\end{figure}

\begin{figure}
\epsscale{0.9}
\plotone{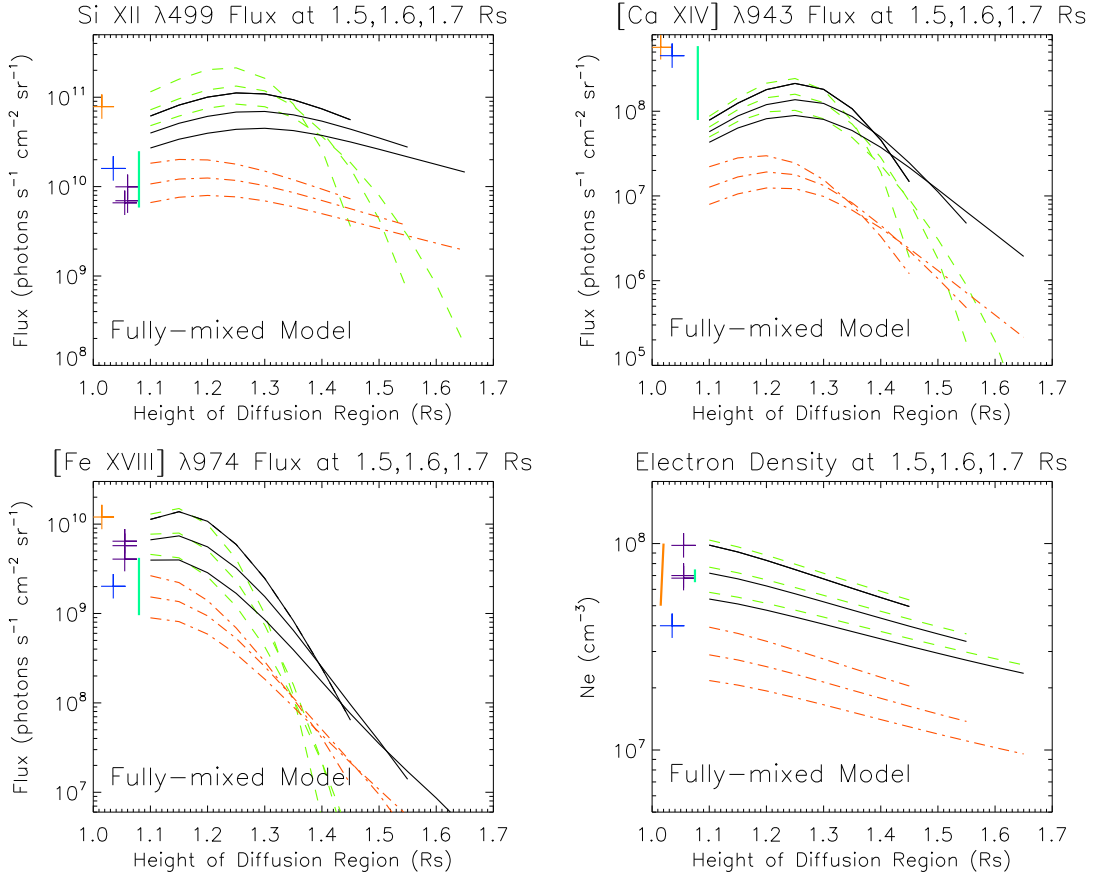}
\caption{For the Fully-mixed model: Line fluxes and electron density observed at 1.5, 1.6, 1.7 \Ro versus the height of the diffusion region for 3 coronal models, N1T1 (green-dash), N1T2 (black-solid), N2T2 (red-dash-dot). Also plotted are the UVCS data (see Fig.10 caption). For the 3 curves corresponding to the 3 observed heights of each model, lower height has higher flux/density in general (cp. Fig.10).  
\label{fig13}}
\end{figure}

\begin{figure}
\epsscale{0.9}
\plotone{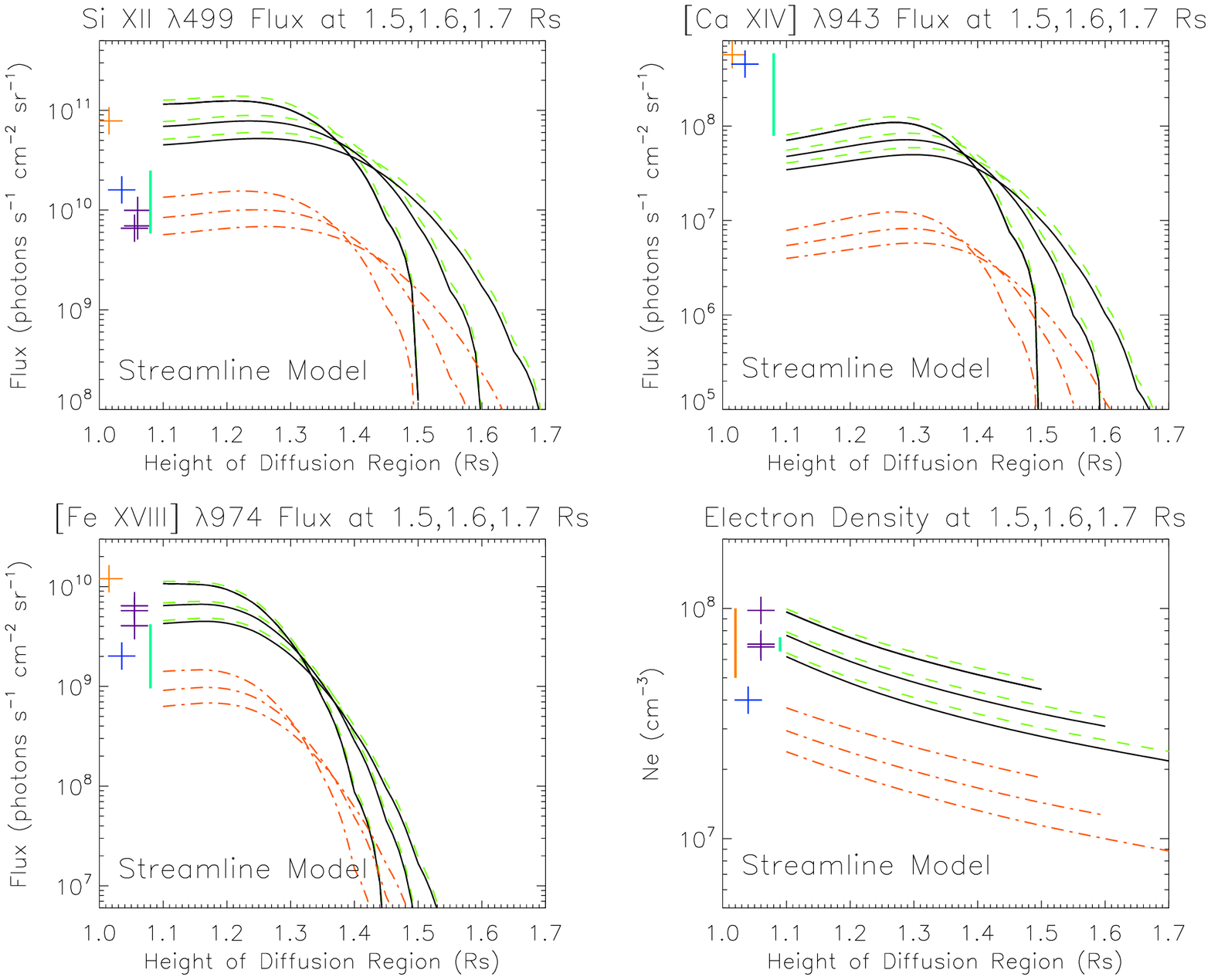}
\caption{Same as Fig.11 but for the Streamline model.
\label{fig14}}
\end{figure}

\clearpage

\begin{figure}
\epsscale{0.7}
\plotone{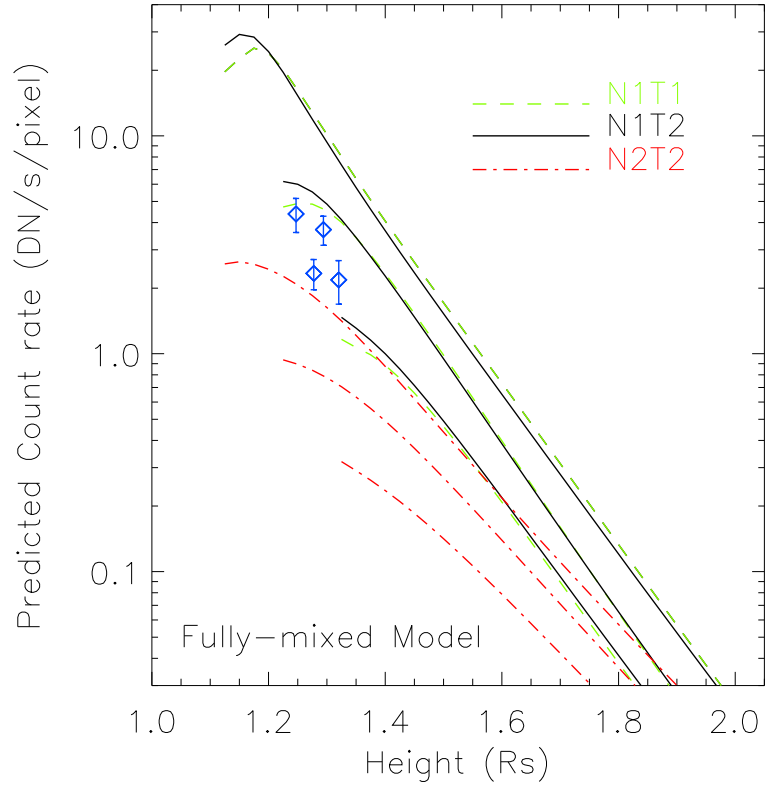}
\caption{Predicted count rate vs. height for Hinode/XRT thin Al-Poly filter for the Fully-mixed model. Plotted are for 3 coronal models, N1T1 (green-dash), N1T2 (black-solid), N2T2 (red-dash-dot), and for 3 DR heights at 1.1, 1.2, 1.3 \Ro.  The LOS depth is assumed to be $5\times 10^3$ km. Also plotted (blue data points) are four measurements of Hinode/XRT on the CS event on Apr.9, 2008 (see Fig.17).
\label{fig15}}
\end{figure}

\begin{figure}
\epsscale{0.7}
\plotone{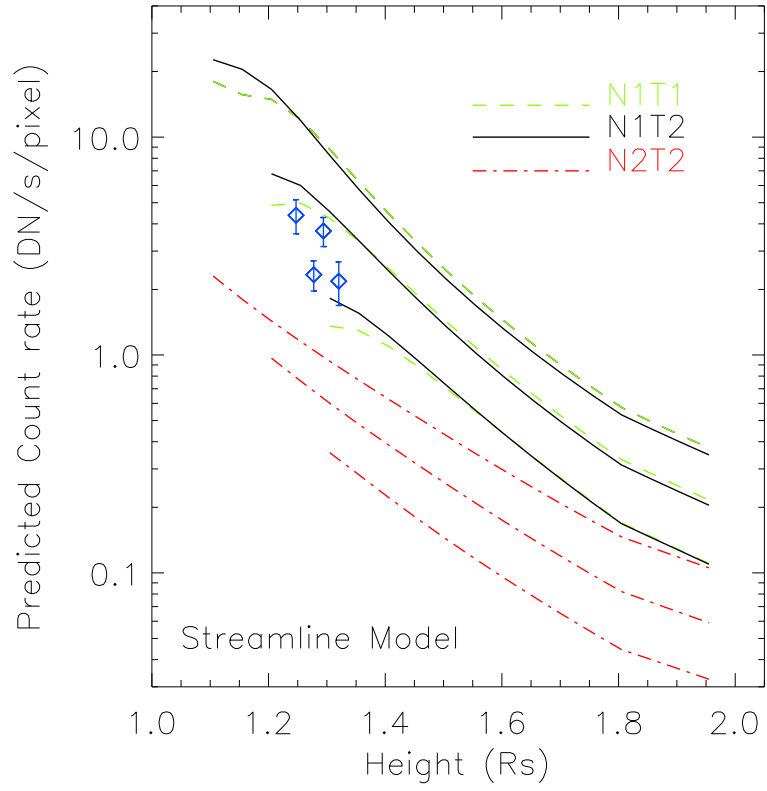}
\caption{Predicted count rate vs. height for Hinode/XRT thin Al-Poly filter for the Streamline model. See caption of Fig.15.
\label{fig16}}
\end{figure}

\begin{figure}
\epsscale{0.85}
\plotone{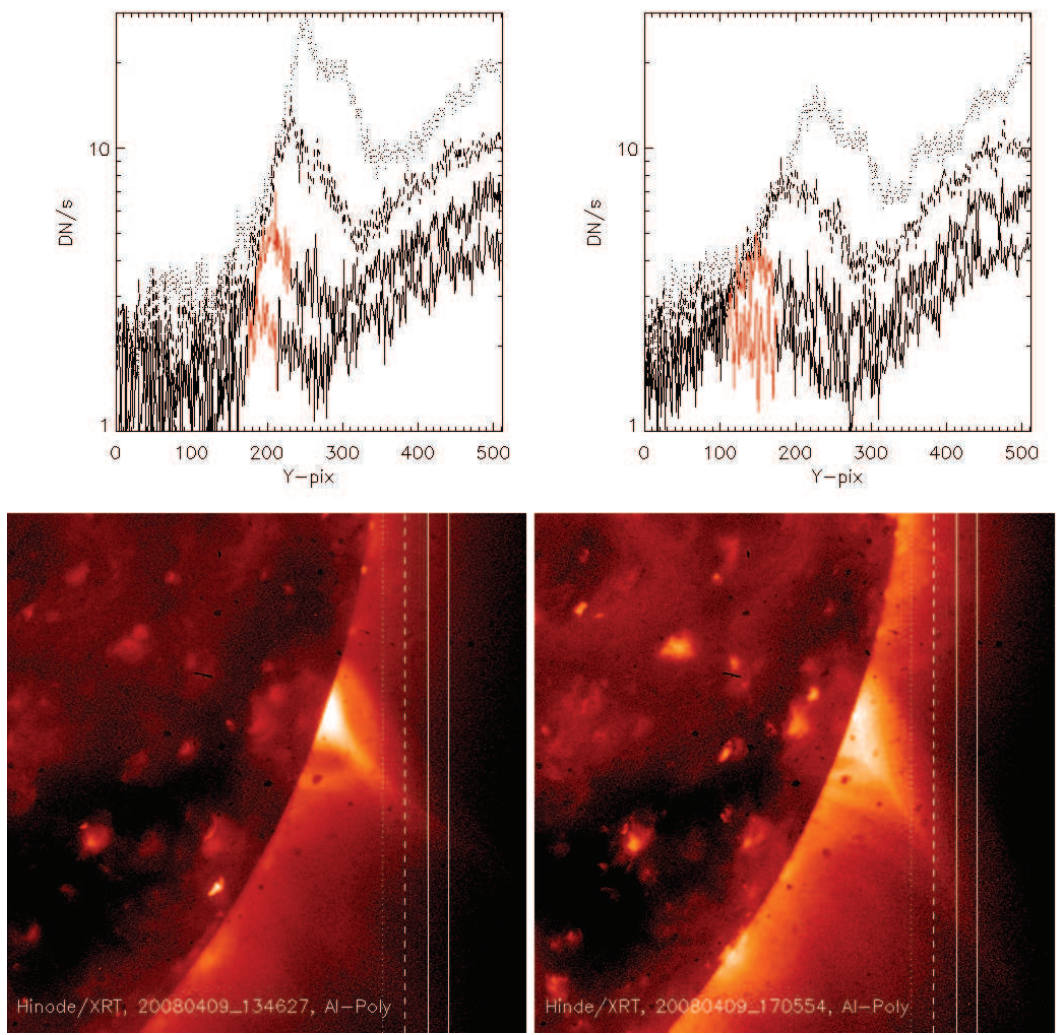}
\caption{Lower panels: Hinode/XRT Al-Poly image on Apr.9, 2008 at 13:46 UT (lower left) and 17:06 UT (lower right) showing the CS extended above the post-CME loops/cusp. The vertical lines are the solar X-positions chosen for the plots in the upper panels. The spatial binning is 1 pixel. Upper panels: observed count rate in DN/s along the Y-pixel number at the 4 solar X-positions marked by the vertical lines in the XRT images. The `bumps' between Y pixel of 100-300 are where the CS is located. Note that, according to Savage et al. (2010), the DR is located in between the dashed and solid lines. Thus our model results, which only apply to the CS/SMS above the DR, should be compared only with the data at the locations marked by the two solid lines (lowest two curves in the upper panels). Plotted in red on the two lowest curves are the Y-pixel ranges used to estimate the mean and standard deviation of the count rate within the CS (shown as blue data points on Figs.15 and 16).  
\label{fig17}}
\end{figure}

\begin{figure}
\epsscale{0.6}
\plotone{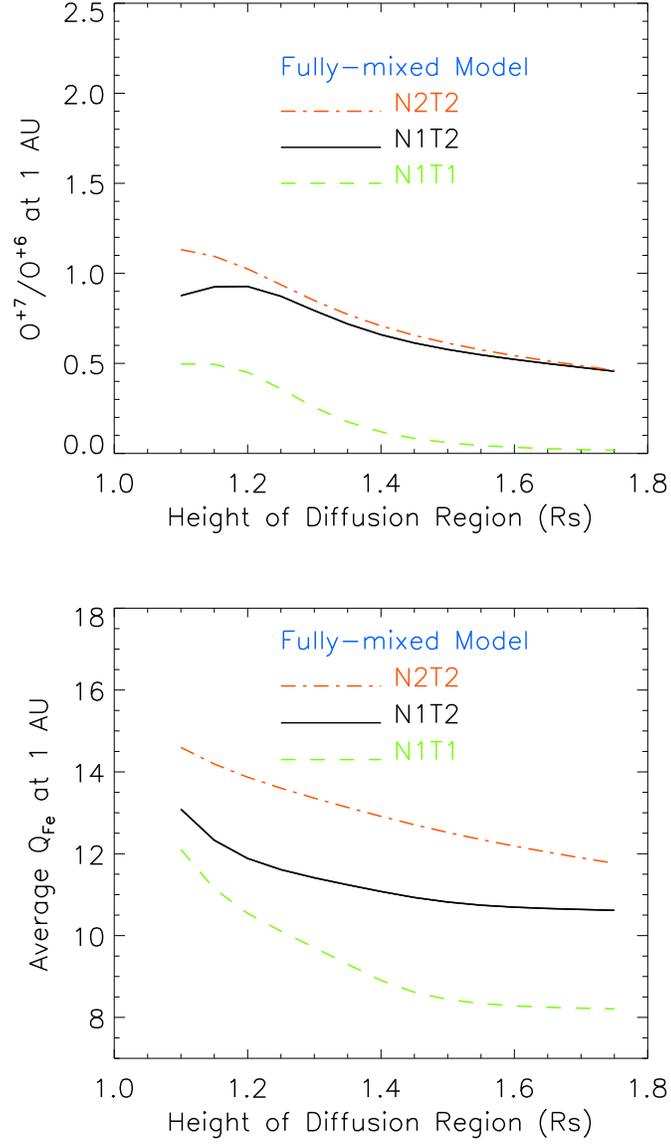}
\caption{O$^{+7}$/O$^{+6}$ ratio and average charge of Fe observed at 1 AU for the Fully-mixed model versus height of the DR with the 3 coronal models compared.
\label{fig18}}
\end{figure}

\begin{figure}
\epsscale{0.6}
\plotone{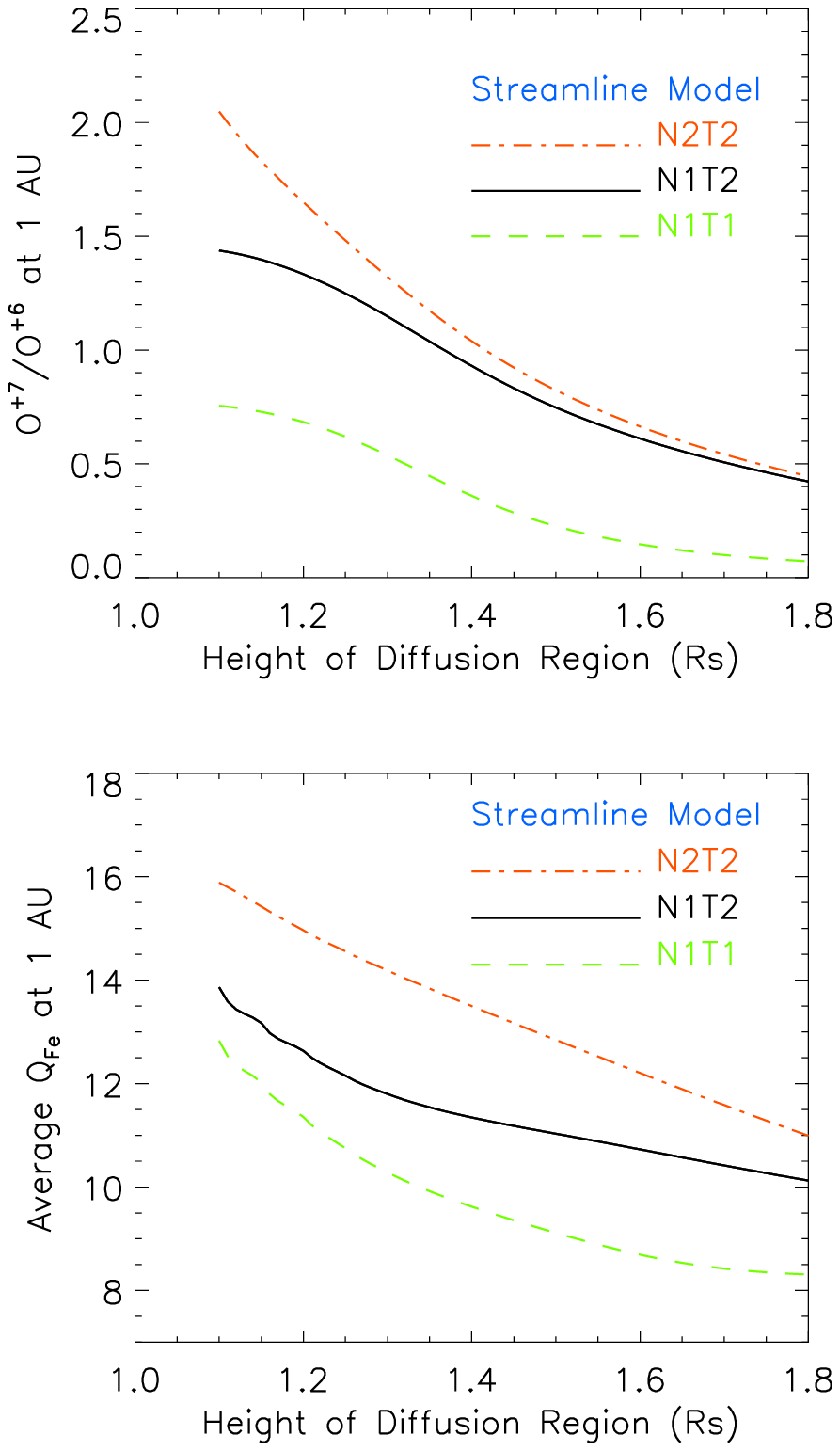}
\caption{Same as Fig.18 but for the Streamline model.
\label{fig19}}
\end{figure}

\clearpage

\end{document}